\documentclass[12pt,preprint]{aastex}
\usepackage{subfigure}
\usepackage[pdftex]{epsfig}

\shorttitle{Infrared Counterparts of Chandra X-ray Sources}
\shortauthors{DeWitt et al.}
 
\begin{document}

\title{Near-infrared counterparts of Chandra X-ray sources toward the
Galactic Center}

\author{Curtis DeWitt\altaffilmark{1}, Reba M.
  Bandyopadhyay\altaffilmark{1}, Stephen S.
  Eikenberry\altaffilmark{1,2}, Robert Blum\altaffilmark{3,5}, Knut
  Olsen\altaffilmark{3}, Kris Sellgren\altaffilmark{4} and Ata
  Sarajedini\altaffilmark{1}}

\email{dewitt@astro.ufl.edu}

\altaffiltext{1}{University of Florida} 
\altaffiltext{2}{University of
  Florida Research Foundation Professor of Astronomy}
\altaffiltext{3}{National Optical Astronomy Observatories, Tucson, AZ
  85719} 
\altaffiltext{4}{Ohio State University}
\altaffiltext{5}{Visiting astronomer, Cerro Tololo Inter-American
  Observatory, National Optical Astronomy Observatory, which are
  operated by the Association of Universities for Research in
  Astronomy, under contract with the National Science Foundation.}
\begin{abstract}

  The \textit{Chandra} X-ray Observatory has now discovered nearly
  10,000 X-ray point sources in the $2^{\circ} \times 0.8^{\circ}$
  region around the Galactic Center \citep{muno09}. The sources are
  likely to be a population of accreting binaries in the Galactic
  Center, but little else is known of their nature.  We obtained
  \textit{$JHK_{s}$} imaging of the $17'~\times 17'$ region around Sgr
  A$^{*}$, an area containing 4339 of these X-ray sources, with the
  ISPI camera on the CTIO 4-m telescope. We cross-correlate the
  \textit{Chandra} and ISPI catalogs to find potential IR counterparts
  to the X-ray sources. The extreme IR source crowding in the field
  means that it is not possible to establish the authenticity of the
  matches with astrometry and photometry alone. We find 2137 IR/X-ray
  astrometrically matched sources: statistically we estimate that our
  catalog contains $289 \pm 13$ true matches to soft X-ray sources and
  $154 \pm 39$ matches to hard X-ray sources. However, the fraction of
  true counterparts to candidate counterparts for hard sources is just
  11 $\%$, compared to 60 $\%$ for soft sources, making hard source
  NIR matches particularly challenging for spectroscopic follow-up. We
  calculate a color-magnitude diagram (CMD) for the matches to hard X-ray
  sources, and find regions where significant numbers of the IR
  matches are real. We use their CMD positions to place limits on the
  absolute \textit{$K_{s}$} band magnitudes of the potential NIR
  counterparts to hard X-ray sources. We find regions of the
  counterpart CMD with $9 \pm 3$ likely Wolf-Rayet/supergiant
  binaries (with 4 spectroscopically confirmed in the literature) as
  well as $44 \pm 13$ candidates that could consist of either main
  sequence high mass X-ray binaries or red giants with an accreting
  compact companion. In order to aid spectroscopic followup we sort
  the candidate counterpart catalog on the basis of IR and X-ray
  properties to determine which source characteristics increase the
  probability of a true match. We find a set of 98 IR matches to
  hard X-ray sources with reddenings consistent with GC distances which
  have a 45\% probability of being true counterparts.

\end{abstract}


\section{Introduction}
The Galactic Center (GC) has been the target of several major
observing campaigns with the \textit{Chandra} X-ray Observatory. One
of the primary results of these campaigns has been to unveil numerous
faint and spectrally hard X-ray point sources toward the GC
region. Hard X-ray sources, with spectral energy distributions (SEDs)
that peak above 1.5 keV, often arise from high extinction. \cite{wang02}
uncovered $\sim$ 1000 such point sources in a $2^{o}\times 0.8^{o}$
field while searching for the source of the He-like Fe emission
emanating from the GC.  To investigate the nature of these point
sources, \cite{muno03a} reobserved the central $17' \times 17'$ around
Sgr A* for an additional 590ks, increasing the total observing time to
626 ks and the number of X-ray point sources to 2357. More than 2000
of the sources were undetected below 1.5 keV, indicating a large amount
of extinction. The line-of-sight to the Galactic Center has an average
column density of $N_{H}=6\times 10^{22}~cm^{-1}$ \citep{baganoff03}, which
causes the flux below 1.5 keV to experience more than $\tau =2$ of
attenuation \citep{tan04}. This suggests that the hard X-ray sources
with attenuated emission below 1.5 keV lie at or beyond the GC
distance.

The faintness and hardness of the X-ray emission from these sources
could in theory be met by background active galactic nuclei (AGN), but
\cite{bandyopadhyay05} find that the number of background AGN within
the $2^{o}\times 0.8^{o}$ field is $< 10\%$, using the source count
distributions from the \textit{Chandra} Deep Field extragalactic
surveys. \cite{muno03a} perform a similar analysis of the number of
background AGNs in the smaller $17' \times 17'$ field and find that
between 20-100 of the 2076 X-ray sources detected at energies above
2keV are AGN, meaning that the bulk of the hard X-ray sources lie
within the Galaxy near to or beyond the GC distance.

\cite{muno03a} enumerate the possible classes to which these
unidentified X-ray sources may belong; these include coronally active
main sequence stars, Young Stellar Objects (YSOs), Algol binaries,
Wolf-Rayet stars (WRs), cataclysmic variables (CVs), pulsars, low mass
X-ray binaries (LMXBs) and high mass X-ray binaries (HMXBs). The soft
foreground X-ray sources could in theory belong to any of these
classes based on their X-ray SEDs; however their low X-ray luminosity
suggests that they may be more likely to be CVs or coronally active
single stars. Of these candidate source types, only LMXBs,
HMXBs and a subclass of CVs with high magnetic fields called
Intermediate Polars (IPs) are known to produce enough hard X-rays to
be detectable through the large extinction to the GC.

\cite{muno03b} search for X-ray flux variability of the sources in the
\cite{muno03a} X-ray point source catalog. They detect 8 sources with
periodic variability among the brightest X-ray sources, but the vast
majority of sources are too faint to search for flux variability at
similar levels.

The next step in identifying the nature of the X-ray sources is to
search for them at other wavelengths. The near-infrared bandpass is
well suited to this effort because stellar photospheres are bright in
these bands and the average extinction is dramatically less than at
optical pass-bands. However, source crowding makes definitive
association between X-ray and IR sources difficult. For example, a $1"$
radius circle at a random location within $10'$ of Sgr A$^{*}$ would
contain a \textit{$K_{s}$}$ < 14.5~ mag$ star more than 25 $\%$ of the
time and this percentage only increases with fainter magnitudes. Thus, the
authenticity of an individual IR/X-ray match can only be verified
through spectroscopy. However, one can also examine the
astrometrically identified candidate counterparts as a group, and make
statistical arguments as to the presence and possible nature of true
counterparts within the catalog of matched sources.

Several groups approach this problem by calculating the excess in the
number of X-ray/IR matches over the number expected by random
coincidence, including \cite{bandyopadhyay05}, \cite{laycock05},
\cite{arendt08}, \cite{gosling09} and
\cite{mauerhan09}. \cite{bandyopadhyay05} used ISAAC on the VLT to
image fields containing 77 of the X-ray point sources discovered by
\cite{wang02}. They detect a small excess in counterparts in the
\textit{J} and \textit{H} bands, but find no significant excess in
matches in the \textit{K} band. However, the positions of a number of
the X-ray sources in the first \textit{Chandra} source catalog were
found be inaccurate at the 0.5$"$ level. A reanalysis of the X-ray/IR
matching of these data sets after the X-ray astrometry was improved
indicates a 2-3 $\sigma$ excess in detections over that expected from
random chance in \textit{J},\textit{H} and \textit{K} bands
\citep{gosling10}.

\cite{laycock05} observe the central $10'\times 10'$ GC area with the
PANIC near-infrared camera on Magellan to a confusion limit of
\textit{$K_{s}$}$=15.4~mag$. Cross-correlating their IR catalog to the
X-ray catalog of \cite{muno03a}, they find a strong IR/X-ray matching
significance for the soft X-ray sources, but little significance for
the hard X-ray sources. They calculate that no more than 10 $\%$ of
the hard X-ray point sources can have apparent \textit{$K_{s}$}
magnitudes brighter than 15 mag. Assuming these hard X-ray sources
reside in the GC, they conclude that certain types of X-ray binaries
that have bright mass-donor stars, such as Be-neutron HMXBs,
must contribute no more than 10 $\%$ to the hard X-ray source numbers.

\cite{arendt08} perform similar matching simulations using a catalog
of $\sim$ 20000 \textit{Spitzer} IRAC point sources detected at
$\lambda=3.6,4.5,5.8$ and $8.0 \mu m$ lying within the $17' \times
17'$ field of the \cite{muno03a} \textit{Chandra} observations. They
found that the soft X-ray sources are correlated to the IRAC catalog,
but they detect no such correlation to the hard X-ray
sources. However, they note that source crowding limits their catalog
to Galactic center sources with $[3.5] < 12.4~mag$, and actual
counterparts may be much fainter than this limit.

Since these studies were published, the Galactic Center X-ray point
source catalog has been updated to include a total of 2 Ms of
accumulated observations of the $2^{\circ} \times 0.8^{\circ}$ area
around Sgr A$^{*}$ \citep{muno09}. 9017 X-ray point sources are
detected, the positions of many of which have been refined from the
earlier catalogs. The majority of the sources in the catalog have
astrometric errors less than 0.7''. \cite{mauerhan09} utilize this
data together with near-infrared data from the SIRIUS camera on the
1.4m telescope at Sutherland Observatory in South Africa, and present
a catalog of 5184 potential near-infrared candidate counterparts to
the X-ray sources.  They find that, statistically, 394 of the NIR
matches to the hard X-ray sources are real counterparts. This amounts
to $5.8 \%$ of the 6760 hard X-ray sources, consistent with the
findings of \cite{laycock05} which limit this percentage to no more
than $10 \%$.

A complete accounting of the fraction of low mass X-ray binaries
(LMXBs) and high mass X-ray binaries (HMXBs) can provide a useful
tracer of both the accumulated star formation and the ongoing star
formation in the GC \citep{grimm03}. The search can also yield rare
and important systems, such as microquasars, X-ray pulsars, and
wind-colliding binaries. \cite{mikles06}, \cite{hyodo08} and
\cite{mauerhan10} have discovered possible wind-colliding O
supergiant and Wolf-Rayet binaries in the GC. These systems could
prove to be important laboratories for studies of high-mass stellar
winds and it may be that more sources of this rare type will be found
amoung the GC X-ray population.

The only way to positively identify individual sources is with
spectroscopy. In the near infrared, active accretion manifests in
emission lines, such as in Br$\gamma$ line emission. Spectroscopic
campaigns such as those described in \cite{mikles06},
\cite{mauerhan10} and \cite{gosling10} have observed only the
brightest infrared matches to X-ray sources.  With the coming advent
of deep multi-object spectroscopy in the near-infrared
\citep{eiken08}, it will be possible to comprehensively follow up
nearly all potential near-infrared counterparts, greatly increasing
the chance of discovering the true nature of these X-ray populations.

In this work, we present new NIR imaging of the deepest part of the
\textit{Chandra} observations in the $17' \times 17'$ area around $Sgr
A^{*}$ containing 4339 X-ray point sources. We create a catalog of
2137 candidate NIR counterparts to X-ray point sources and
statistically identify the locations of the true counterparts on a NIR
color-magnitude diagram. We also isolate the characteristics that make
likely authentic counterparts easier to find amongst the large number
of spurious matches. These results will greatly aid the target
selection for future spectroscopic observations as well as indicating
the likely types of interacting binaries we may identify within this
source population.

\section{Observational Data}

We use four sets of data in this work: the \textit{Chandra} X-ray
point source catalog of \cite{muno09}, a near-infrared \textit{J},\textit{H} and
\textit{$K_{s}$} catalog from our observations with ISPI at the CTIO 4m
telescope, the 2MASS \textit{$JHK_{s}$} all sky data release
\citep{tmass}, and the \textit{JHK} imaging data and catalogs of the
UKIDSS Galactic Plane Survey \citep{lucas08}. The 2MASS and UKIDSS
data were used to photometrically and astrometrically
calibrate the ISPI data, while the ISPI data and \textit{Chandra} data
were used to create a crossmatched source catalog.
	
\subsection{X-ray} 
\cite{muno09} compile 88 \textit{Chandra} observations of the Galactic
Center (GC) taken between April 2000 and August 2007 and find 9017
point sources over a $2^{\circ}\times 0.8^{\circ}$ field. The $17'
\times 17'$ field centered on Sgr A$^{*}$ has been the target of 1Ms
of exposures and has some of the best sensitivity ($4 \times 10^{31}$
ergs s$^{-1}$ for GC sources at 8 kpc distance) and the best
positional accuracy in the X-ray catalog.  76 $\%$ of the X-ray
sources in this region have positional accuracies of 0.7$''$ and
better.
\subsection{ISPI} 
We used the ISPI camera on the CTIO 4m Blanco Telescope to observe the
$17'\times 17'$ central region of the GC, covering the region of
highest source density and in sensitivity in the \textit{Chandra} X-ray
data. ISPI has a $10.5'$ square field of view and a $0.3''$/pixel
plate scale \citep{vdb04}.  We observed the $17' \times 17'$ region
with the \textit{J}, \textit{H} and \textit{$K_{s}$} filters, on
August 10, 2005.

\begin{figure}[ht]
\centering
\subfigure[] 
{
    \label{fig:color:a}
    \includegraphics[width=7cm]{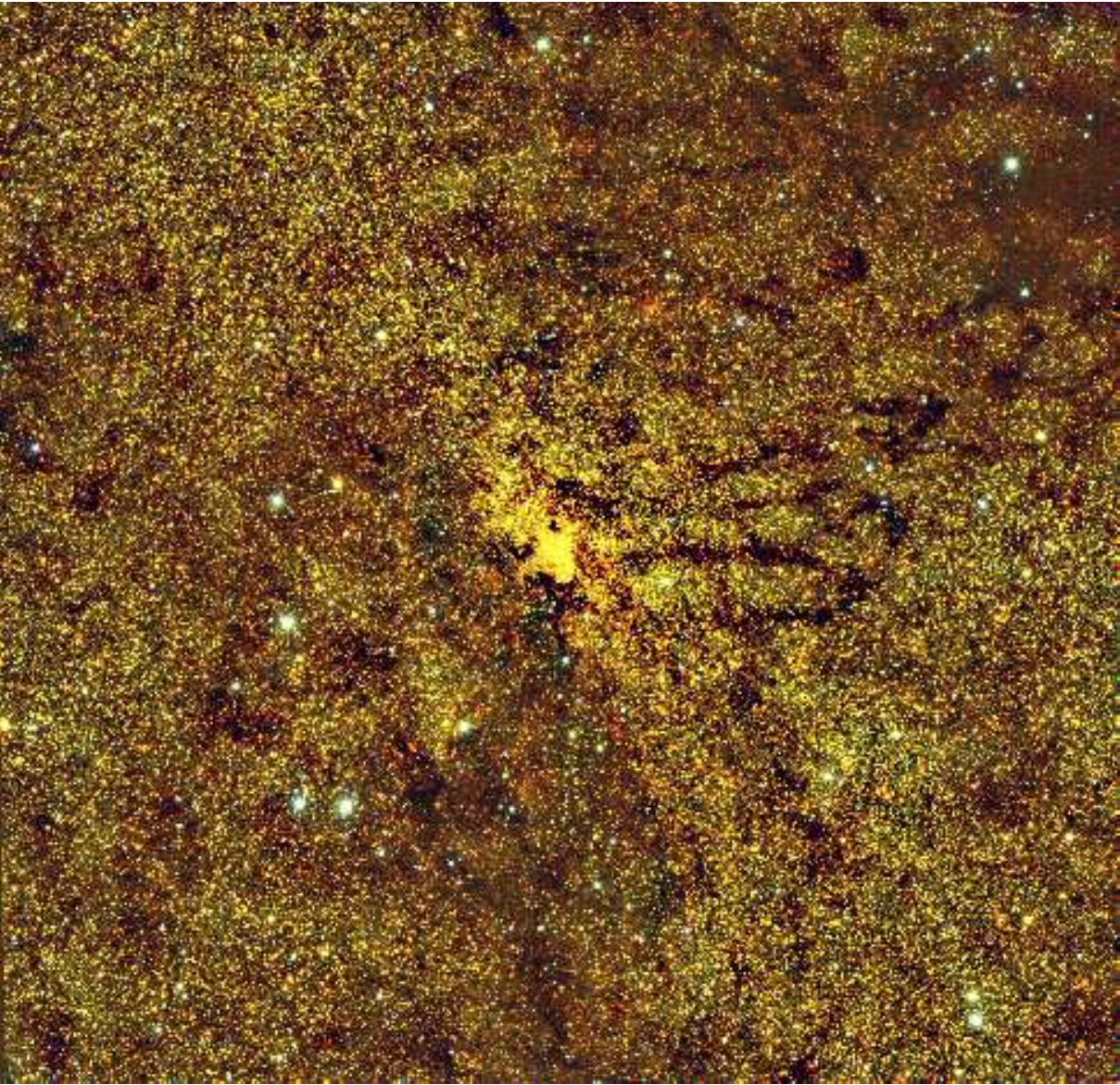}
}
\hspace{0.1cm}
\subfigure[] 
{
    \label{fig:colorimage:b}
    \includegraphics[width=7cm]{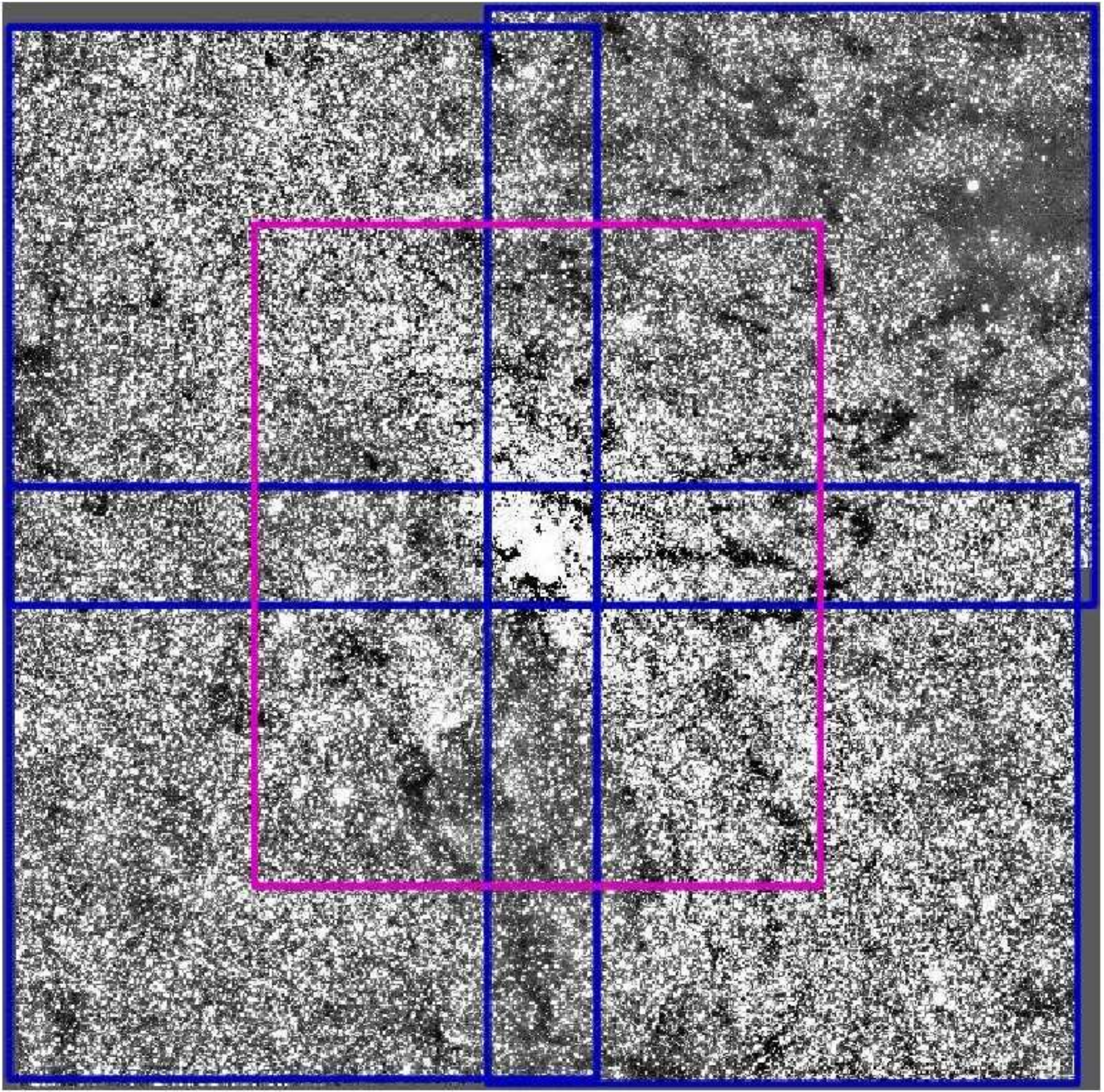}
}
\caption{(left) 3 color $JHK_{s}$ composite image of ISPI Galactic
  center data. Each side is 17' in length. North is up and east is to
  the left.  (right) The pointing pattern of the ISPI observations,
  overlaid on the \textit{$K_{s}$} image. The magenta box represents
  the $10.5' \times 10.5'$ field of view of a single pointing with
  ISPI. The full frame is a $17' \times 17'$ region and consists of 5
  such pointings. }
\label{fig:colorimage} 
\end{figure}

We obtained IR imaging of the full \textit{Chandra} field of \cite{muno03a}
using a total of 4 pointings. The individual exposure times were
short, 3.2s for \textit{H} and \textit{$K_{s}$} and 5s for \textit{J},
so that the number of saturated bright sources would be kept to a
minimum. We used a 5 point dither pattern with a nod length of $20"$
to compensate for the detector's cosmetic defects. The total exposure
times for each quadrant of the field were 200s, 113s and 32s for
\textit{J}, \textit{H} and \textit{$K_{s}$}, respectively. Between
pointings we checked the image focus by observing the center of the
field. The focus stability was very good, which allowed us to use
these frames for the creation of a fifth pointing, overlapping the
other 4 pointings. This amounted to obtaining an additional frame with
exposure times of 60s, 68s and 68s for the \textit{J}, \textit{H} and
\textit{$K_{s}$} filters. The typical seeing for all images was $\sim
0.9''$.  The theoretical limiting magnitudes (not including crowding
effects) for the fields were \textit{J}$=19.5~mag$,
\textit{H}$=18.5~mag$, and \textit{$K_{s}$}$=17.7~mag$, taken from the
CTIO exposure calculator for ISPI. For sky background removal we
observed an offsource field 10 degrees south of the GC, with the same
exposure times and the same 5 point dither pattern in all 3
filters. Dome flats and dark frames were taken at the beginning and
end of the night. The left panel of Figure \ref{fig:colorimage} shows
a 3-color composite image of the full $17' \times 17'$ field and
demonstrates the extreme stellar crowding in this region. The right
panel of Figure \ref{fig:colorimage} displays the pointing pattern of
our observations.

\subsection{2MASS data}

The 2MASS catalog is an all-sky \textit{J}, \textit{H} and
\textit{$K_{s}$} survey that has become a standard infrared photometry
reference \citep{tmass}. The astrometric precision of 2MASS is
reported as $\sim 0.15''$ RMS. We used this catalog for calculating
the ISPI image distortion and defining the WCS of the images. However
the depth of the survey is significantly shallower than the ISPI data,
with $10\sigma$ limits of \textit{J}$=15.8~mag$, \textit{H}$=15.1~mag$
and \textit{$K_{s}$}$=14.3~mag$ and a point spread function (PSF) full
width half maximum (FWHM) which was typically between $2.5"$ and
$3"$. The severe crowding in the GC makes the ``effective'' magnitude
limits even brighter than these nominal values. The combination of the
high crowding and larger PSF of 2MASS meant that many stars that could
have been used as photometric references for our ISPI field were badly
blended in 2MASS. 

\subsection{UKIDSS Data}
The UKIRT Infrared Deep Sky Survey (UKIDSS) is a set of 5 NIR imaging
surveys with complementary goals \citep{lawrence07}.  One of the
subsurveys is the Galactic Plane Survey (GPS), which includes
\textit{JHK} coverage of the region between $-5^{\circ} < l <
15^{\circ}$ and $|b| < 2^{\circ}$ \citep{lucas08}.  The nominal survey
depth was \textit{J}$=19.9~mag$, \textit{H}$=19.0~mag$ and
\textit{K}$=18.8~mag$. The plate scale of the UKIRT IR camera is
$0.4"$, but using the technique of microstepping the delivered plate
scale of the images is $0.2"$. The typical seeing is $0.9"$ for the
data covering the region of our ISPI field. The primary data product
of the survey is a source catalog produced with an aperture photometry
pipeline. We rederived the photometry of UKIDSS GC data with
PSF-fitting in DAOPHOT, in part to better match the PSF photometry
that we performed with ISPI and in part because the aperture pipeline
catalogs did not include some of the highest stellar density regions
in the field.  The details of the UKIDSS PSF reduction followed the
same procedures as for ISPI PSF reduction, described in \S 3.1, and
will be described in full in a subsequent paper (DeWitt et al., in
prep.).  The \textit{JHK} filter set for the UKIDSS survey is slightly
different from the \textit{$JHK_{s}$} set used by 2MASS and ISPI. We
converted the UKIDSS \textit{JHK} magnitudes to the 2MASS system using
the prescription found in \cite{hodgekin09}. The depth and resolution
of the UKIDSS data are a closer match to the ISPI data than 2MASS,
allowing better star-to-star comparison of the photometry. We used our
PSF-fit UKIDSS catalog along with the 2MASS catalog for photometric
calibration of the ISPI data.

 \section{Data Reduction}
\subsection{Creating the ISPI catalog}
The ISPI images were reduced with the FATBOY pipeline, an imaging and
spectroscopy package developed at University of Florida
\citep{warner10}. FATBOY uses Python routines similar to the standard
image reduction tasks found in IRAF. The processing steps include
pixel linearization, dark subtraction, flat-field correction,
cosmic-ray removal, sky-background subtraction and image stacking. We
made a 6$^{th}$ order distortion correction for each color using the
2MASS data, and used it to stack images in each color and pointing into an
image of uniform plate scale. At this stage we had 5 images in each
filter covering the full $17' \times 17'$ field. 

The surface density of detected point sources was very large: 0.03,
0.25 and 0.25 sources$/$arcsec$^{2}$ for \textit{J},\textit{H} and
\textit{$K_{s}$}, respectively. These densities were unsuitably large
for standard aperture photometry, and we therefore employed DAOPHOT
and ALLSTAR, which are specially designed for crowded fields
\citep{stetson87}, to perform PSF-fitting photometry on each of the
stacked images.

We locked the astrometry to the 2MASS catalog by using the IRAF task
\textit{MSCTPEAK}. The residual RMS error of the ISPI plate solution
with respect to \textit{2MASS} was $0.2"$. The internal astrometric
precision of 2MASS is reported to be 0.15'' \citep{tmass}.
We used the 2MASS catalog to find photometric zeropoints for each
image, by positionally matching the bright, unblended 2MASS sources to
stars found in the 2MASS catalog and then calculating the median
offset.  At this point we had five catalogs for each filter, one for
each ISPI pointing, with significant overlap in coverage between the
catalogs.

We merged the per-pointing ISPI catalogs of each filter using a $0.3"$
error radius. We combined the catalog entries by either averaging the
photometry, or in the case where the same star differed in brightness
by more than the 1$\sigma$ error for that magnitude (see \S 3.2), we
adopted whichever brightness measurement had the lower PSF fitting
residuals reported by \textit{DAOPHOT}. The result was a single
catalog for each of the \textit{J}, \textit{H} and \textit{$K_{s}$}
filters, covering the entire $17' \times 17'$ ISPI GC field. 

We merged the \textit{J}, \textit{H} and \textit{K$_{s}$} catalogs
into a single three color catalog by matching sources between the
individual filters with a 0.3$"$ error radius. The standard deviation
of the positions of sources matched across filters is $\sim 0.1"$,
which means our $0.3"$ match threshhold should match sources with up
to 3$\sigma$ positional errors. First the $J$, $H$ and $K_{s}$
catalogs were matched in pairs, i.e. $JH$,$JK_{s}$ and $HK_{s}$. To
locate the 3-filter matches we identified entries from the $JH$
catalog and $HK_{s}$ catalog with the same H band source. Fewer than
$0.5\%$ of sources had multiple astrometric matches in another band.
In the few cases when there was a second astrometric match, we adopted
the closer match for our final merged catalog.  We used the $K_{s}$
band position whenever available for the final catalog astrometry. If
a source was not detected in $K_{s}$, we used the $H$ position, and if
$H$ was unavailable we used the $J$ position.

The final catalog contains 241,552 distinct sources, consisting of
50,851 detections in \textit{J}, 200,302 detections in \textit{H} and
192,292 detections in \textit{K$_{s}$}. The reason for the greater
number of detections in \textit{H} than in \textit{K$_{s}$} is mostly
due to $\sim 24,000$ arcsec$^{2}$ of missing coverage in the
\textit{$K_{s}$} band on the outer parts of the observed fields. These
areas were observed in \textit{$K_{s}$} as part of the dither pattern
of one of the pointings, but they were discarded because two
\textit{$K_{s}$} images were blurred by telescope shake. This region
with only \textit{J} and \textit{H} band coverage lies on the outer
edges of the North-East quadrant of \ref{fig:colorimage} and makes up
just $\sim 2\%$ of the total $17' \times 17'$ area of the ISPI GC
field.

\subsection{Photometric errors and completeness} 

\begin{figure}[!h]
\centering
\subfigure[] 
{
    \includegraphics[width=7cm]{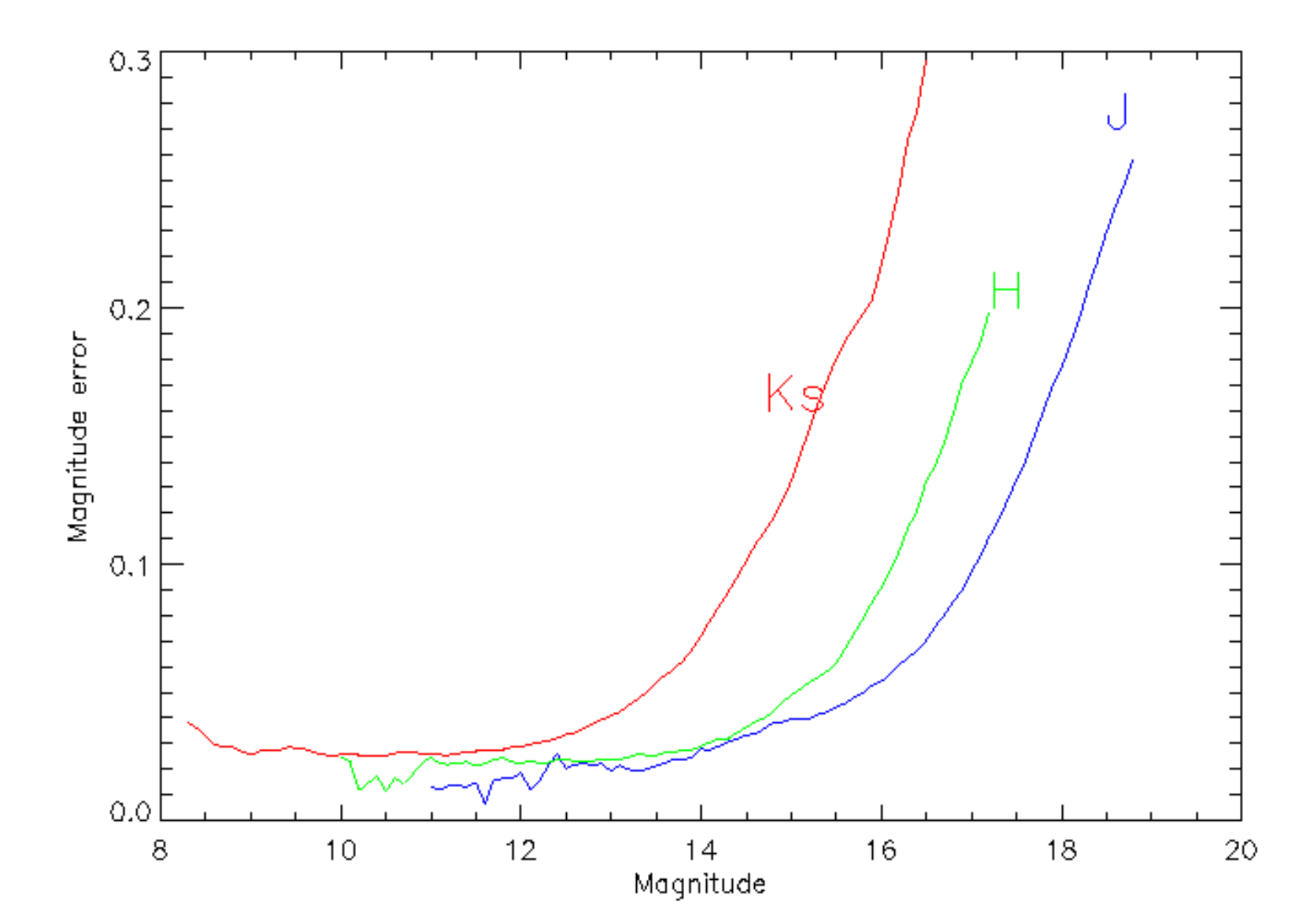}
}
\hspace{0.1cm}
\subfigure[] 
{
    \includegraphics[width=7cm]{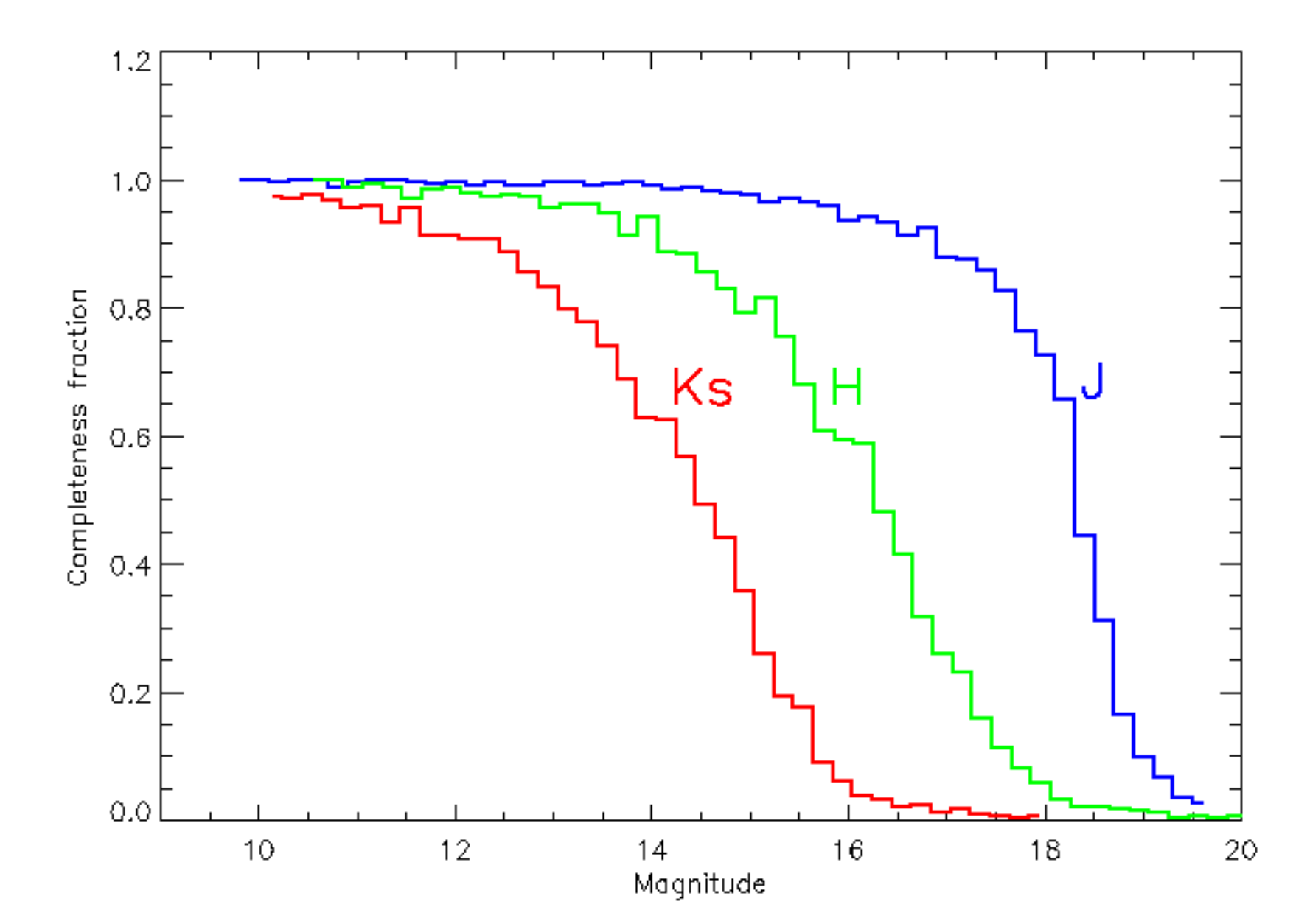}
}
\caption{(left) ISPI Photometric Error: Error per filter derived using
  multiple measurements of the same stars. (right) ISPI photometric
  completeness: The fraction of artificial stars recovered at a given
  $J$,$H$ or $K_{s}$ magnitude.}
\label{fig:errandcomp} 
\end{figure}

To derive photometric errors we used regions of the field that were
observed multiple times in the ISPI data. We calculated the
differences in the measured magnitudes, and took the standard
deviation of all stars in a certain magnitude bin to represent $\sqrt
2 \times$ the error for that magnitude. This error should encompass
photon noise and errors associated with PSF-fitting, but it will be an
underestimate when stellar blending dominates the noise. The error
versus magnitude relation for each filter is shown in the left panel
of Figure \ref{fig:errandcomp}.

We calculated the completeness fraction of our catalogs by injecting
artificial stars of various magnitudes into the images with the
DAOPHOT tool \textit{addstar}. The stars were overlaid on the original
images in a sparse rectangular grid where the magnitudes and x and y
values of the grid position were randomly assigned. We made sure that
the magnitude range covered the interval where between $1\%$ and
$95\%$ of stars were recovered. We restricted the number of injected
stars to fewer than $10~\%$ of the number of stars already contained
in the image in order to prevent the introduction of different
crowding characteristics. The \textit{DAOPHOT} and \textit{ALLSTAR}
parameters were kept the same as in the original reduction. We counted
an artificial star as recovered if it was found within $0.5"$ of its
injected position and we excluded any artificial stars that fell
within $0.5"$ of a real star in the ISPI images. Figure
\ref{fig:errandcomp} shows the completeness fraction versus magnitude
for artificial stars for the $J$, $H$, and $K_{s}$ filters. These
values should reflect typical regions in the ISPI field, but the local
completeness limit will likely vary with the source density changes
across the field.

In Table \ref{tab:catalogcharacteristics} we show the number of
sources in the final merged ISPI catalog, the $5\sigma$ limiting
magnitude predicted by the ISPI exposure calculator from Poisson
statistics, the $5\sigma$ magnitude limit (where the magnitude error
is $\sim 0.2~mag$) calculated from the photometry of stars observed
multiple times, and the mean $50~\%$ completeness magnitude for each
of the filters.
\begin{table}[ht]
\begin{center}
\caption{ISPI point source catalog characteristics.}
\label{tab:catalogcharacteristics} 
\begin{tabular}{crrrrrrrrrrr}
  \tableline\tableline
  Filter & Number of stars &  5-$\sigma$ Mag. limit, & 5-$\sigma$ Mag. limit,& $50 \%$ Completeness\\
  &                 &  Poisson error & fitting error &     limit \\
  \tableline
  $J$       &  50851 & 19.5 & 18.3 &18.4 \\
  $H$       & 200302 & 18.7 & 17.2 &16.4 \\
  $K_{s}$   & 192292 & 17.3 & 15.8 &14.5 \\
  \tableline
\end{tabular}
\end{center}
\end{table}

\section{Real and spurious IR matches to the Chandra X-ray catalog}

There are 4339 X-ray point sources from the \cite{muno09} catalog that
lie within our field. We exclude any X-ray sources with $\sigma_{X} >
2.0"$, where $\sigma_{X}$ is the $95\%$ positional uncertainty for the
X-ray sources, in arcseconds. The 71 sources with larger errors have a
median number of possible matches of 5 each, which would preclude
useful follow-up. For the rest of this work we only consider the
remaining 4268 X-ray sources.

We cross-correlate these 4268 X-ray sources with the ISPI catalog
using a $\sigma_{X} + 0.2"$ matching radius, where $0.2"$ represents
the IR catalog astrometric error. We do not use the quadrature sum of
these errors, $\sqrt{\sigma_{X}^{2} + 0.2^{2}"}$, because it
underestimates our positional uncertainty. The imperfect distortion
correction contributes a small, regional systematic error to the
$0.2"$, which we discovered during the filter merging stage of the
ISPI catalog creation. This type of error should have an additive
effect on the separation between X-ray and NIR sources. This matching
criterion is imperfect, and may exclude a small number of actual
NIR/X-ray matches with $\sim 3\sigma$ positional errors, but we have
compromised in order to obtain fewer spurious matches so that we can
extract useful information from the matched sources.

With this matching criterion in effect, we find that $43.2 \%$ of
X-ray sources have 1 or more candidate NIR counterparts within their
matching radius, and $5.5\%$ have 2 or more counterparts
(see Table \ref{tab:nmatch}). Our catalog of ISPI/X-ray matches is available
online as a searchable database called \textit{Chandra Galactica}. The
website url is http://galcent.astro.ufl.edu. A sample of the catalog
is shown in Table 9.

\begin{table}
\begin{center}
  \caption{Number of X-ray sources with 0,1,2, or $>$ 3 astrometric
    matches in the ISPI $JHK_{s}$ catalog.}
\label{tab:nmatch} 
\begin{tabular}{crrrrrrrrrrr}
\tableline\tableline
Number of Matches & Occurences & Percentage\\
\tableline
0       & 2425 & $56.8\%$ \\
1       & 1610 & $37.7\%$ \\
2       & 192  & $4.5\%$  \\
$>$3    & 41   & $1.0\%$  \\
\tableline
\end{tabular}
\end{center}
\end{table}

In Figure \ref{fig:maghists} we show magnitude and color histograms
for all the ISPI sources and the candidate matches to X-ray
sources. The histogram for the X-ray candidate counterpart IR sources
is overplotted in red and normalized to the total number of sources in
the whole ISPI catalog. Inset in each histogram is the same data
plotted as a cumulative distribution. The \textit{J} band clearly
shows a difference between the candidate counterparts and the field
source populations. The fact that the distribution of the \textit{J}
band magnitudes of candidate counterpart is distinct from the entire
catalog suggests that at least some of these matches are real. The
\textit{H} and \textit{K$_{s}$} candidate counterpart magnitude
distributions are less distinct from the overall ISPI catalog, which
is probably due to the higher rate of spurious matches in these bands.

The $(J-H)$,$(J-K_{s})$ and $(H-K_{s})$ color histograms in the lower
panels of Figure 3 all clearly show an excess in
matches to less reddened foreground sources. This suggests that a
significant number of the unreddened candidate counterparts to X-ray
sources are authentic counterparts.

In Figures \ref{fig:cmd} and \ref{fig:cc} we display a
\textit{H-K$_{s}$}/\textit{K$_{s}$} CMD and
a \textit{J-H}/\textit{H-K$_{s}$} color-color diagram of the matched
NIR sources and the entire ISPI catalog. The CMD clearly shows 2
populations of sources: sources with low reddening in a nearly
vertical line at \textit{H-K$_{s}$}$=0.2~mag$ and sources with high
reddening centered at \textit{H-K$_{s}$}$=2.0~mag$. The sources with
low reddening are nearby field sources, while the sources with
high reddening should be near the GC. The color-color diagram shows
that the sources are dispersed along the reddening vector. By eye, the
X-ray source matches marked in color on these figures show little
difference from the overall ISPI catalog. This would be expected
either if the sources actually are not distinct from the overall NIR
source population or if most of the matches are spurious. To evaluate
the number and characteristics of the probable true matches, we need
to have an idea of the number of matches that would happen simply by
chance.

\begin{figure}[ht]
\centering
\label{fig:maghists}
\subfigure[$J$] 
{
    \label{fig:colordistributions:a}
    \includegraphics[width=5cm]{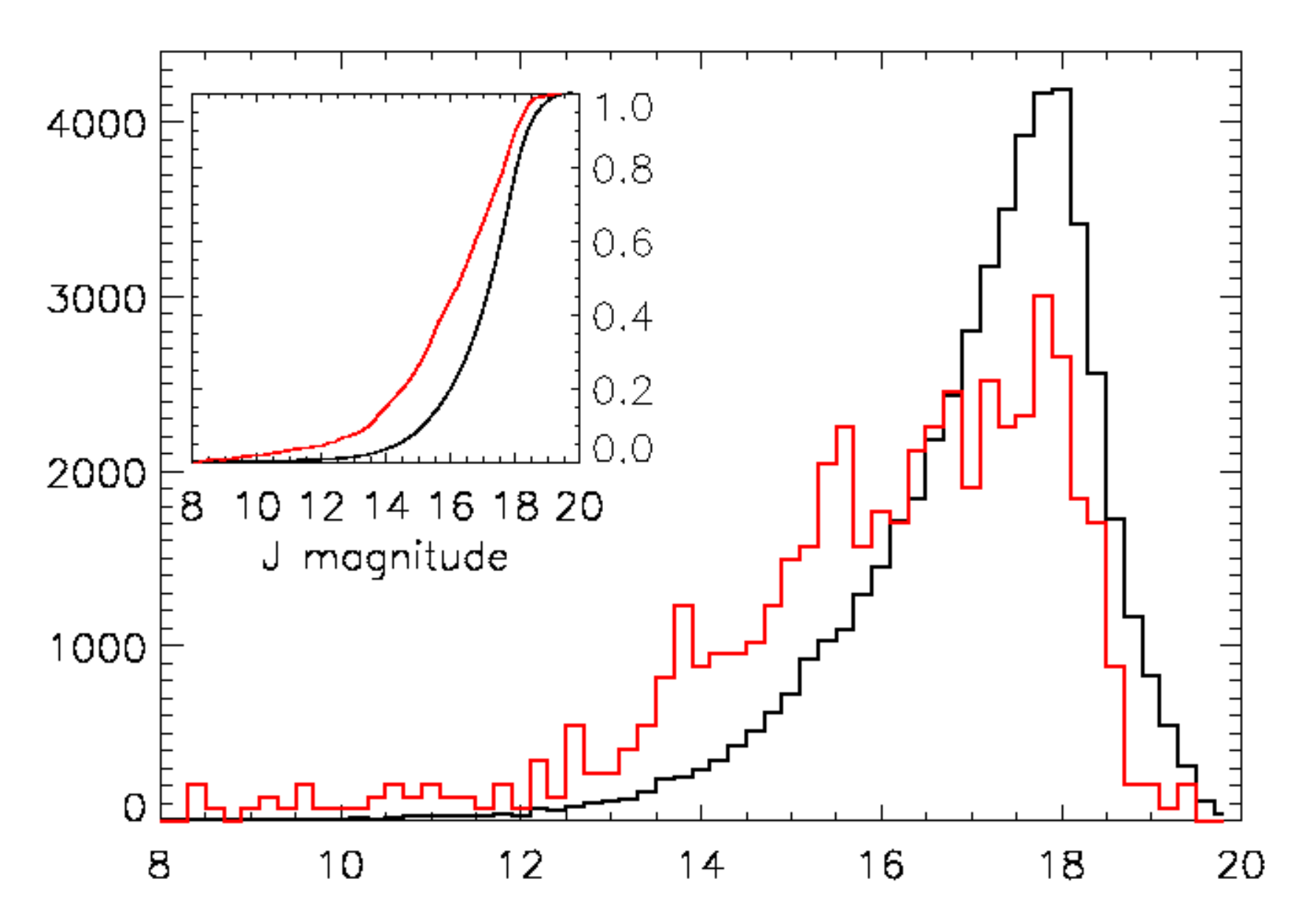}
}
\hspace{-0.6cm}
\subfigure[$H$] 
{
    \label{fig:colordistributions:b}
    \includegraphics[width=5cm]{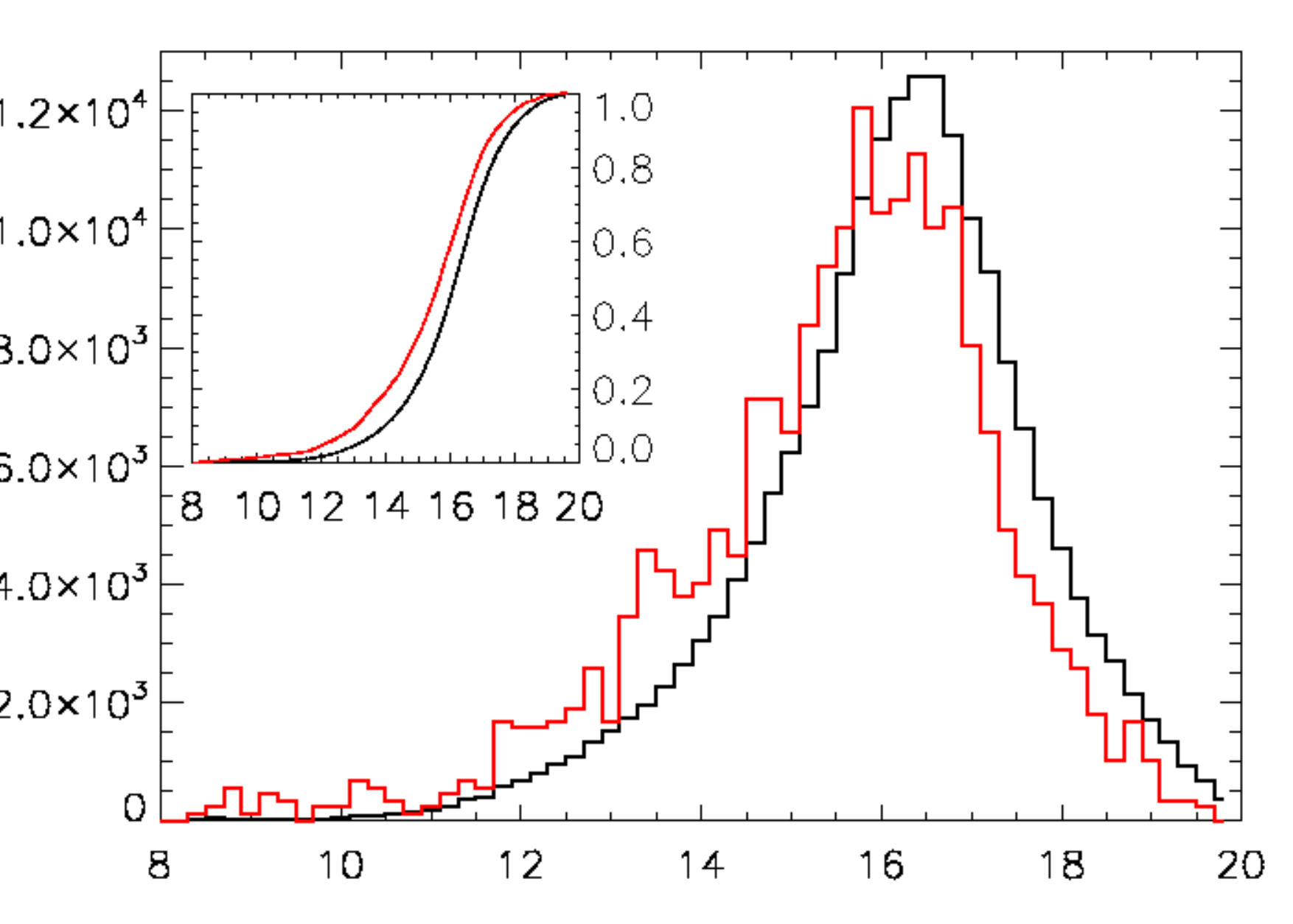}
}
\hspace{-0.6cm}
\subfigure[$K_{s}$] 
{
    \label{fig:colordistributions:c}
    \includegraphics[width=5cm]{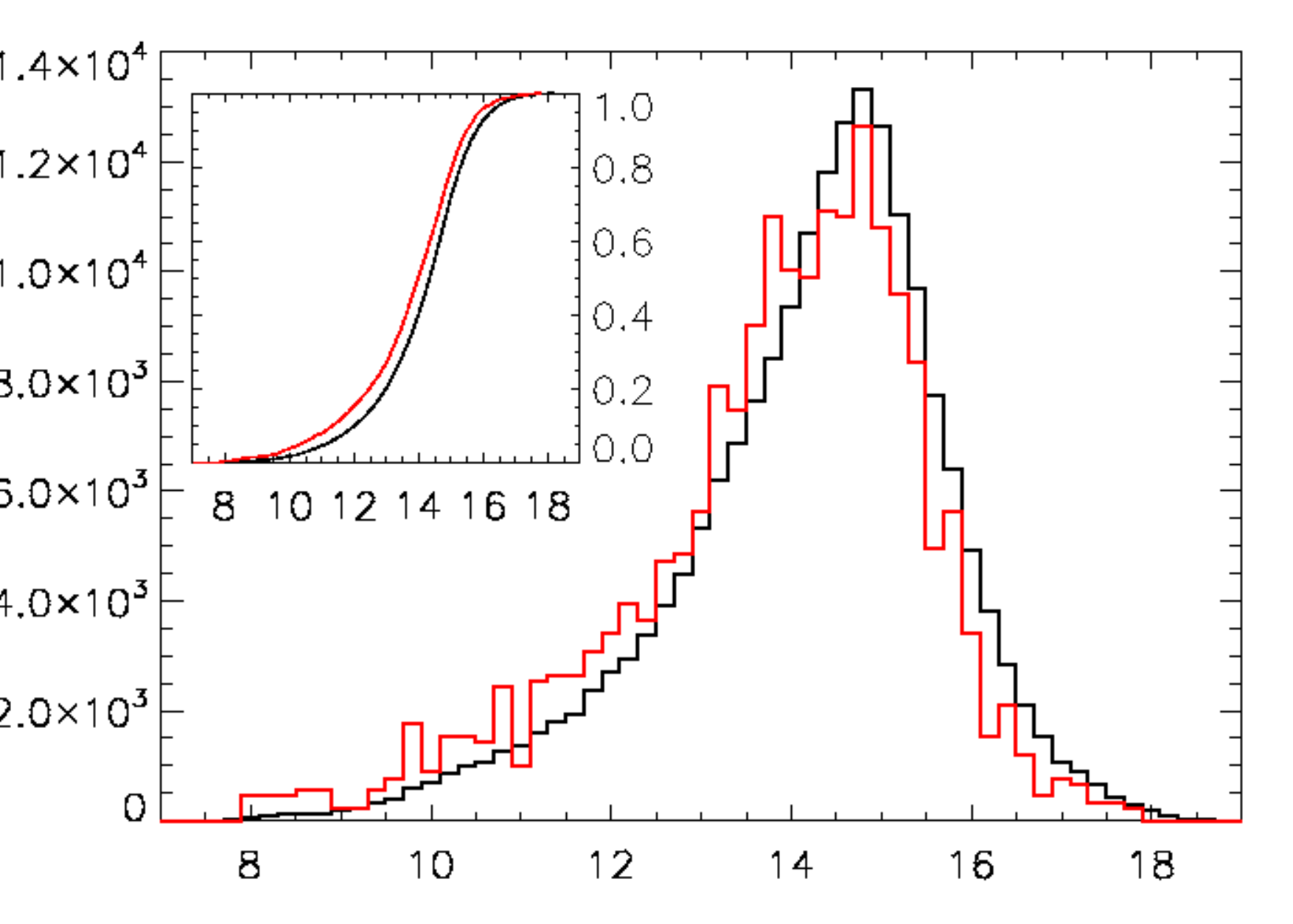}
}
\hspace{-0.6cm}
\subfigure[$(J-H)$] 
{
    \label{fig:colordistributions:d}
    \includegraphics[width=5cm]{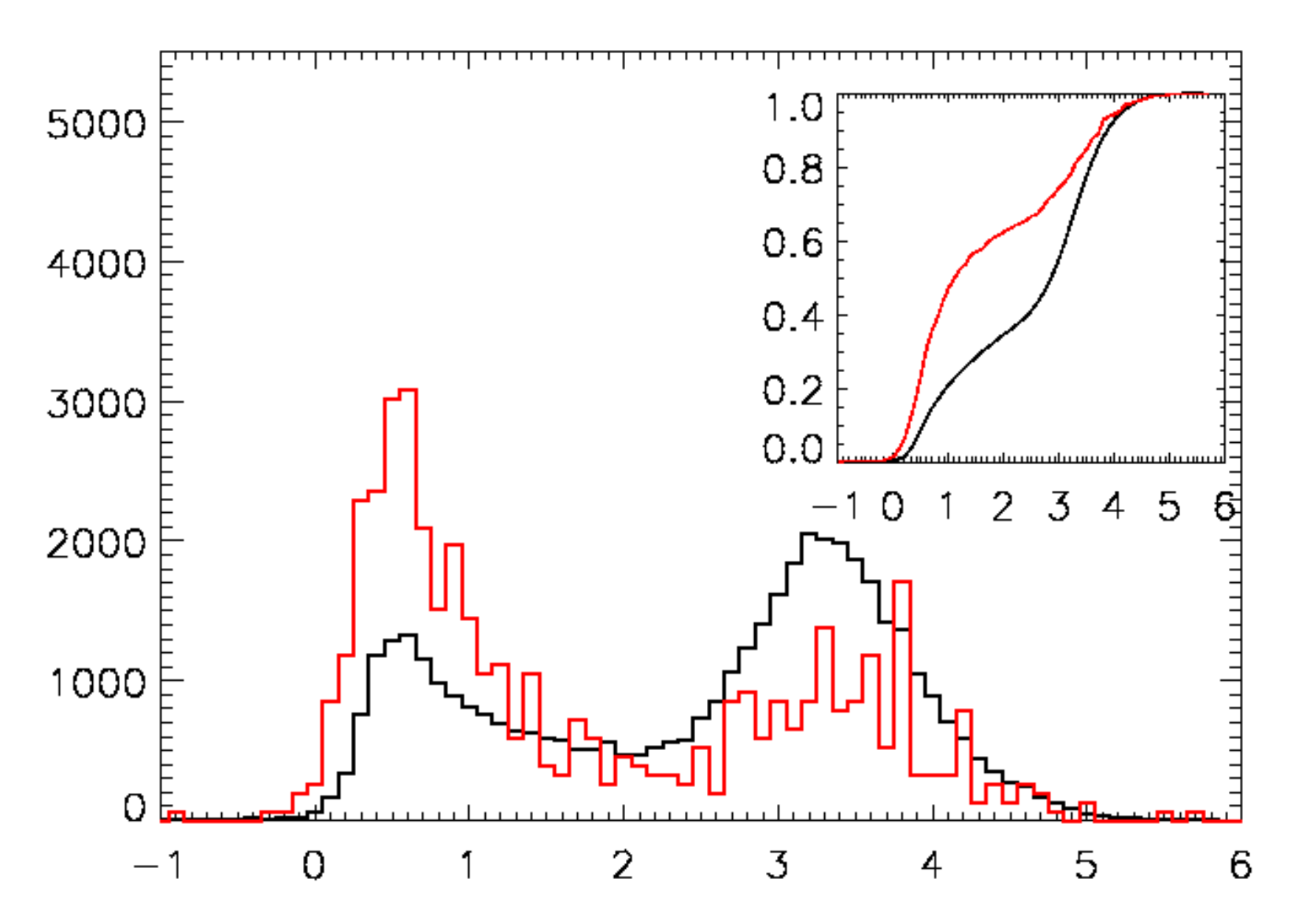}
}
\hspace{-0.6cm}
\subfigure[$(J-K_{s})$] 
{
    \label{fig:colordistributions:e}
    \includegraphics[width=5cm]{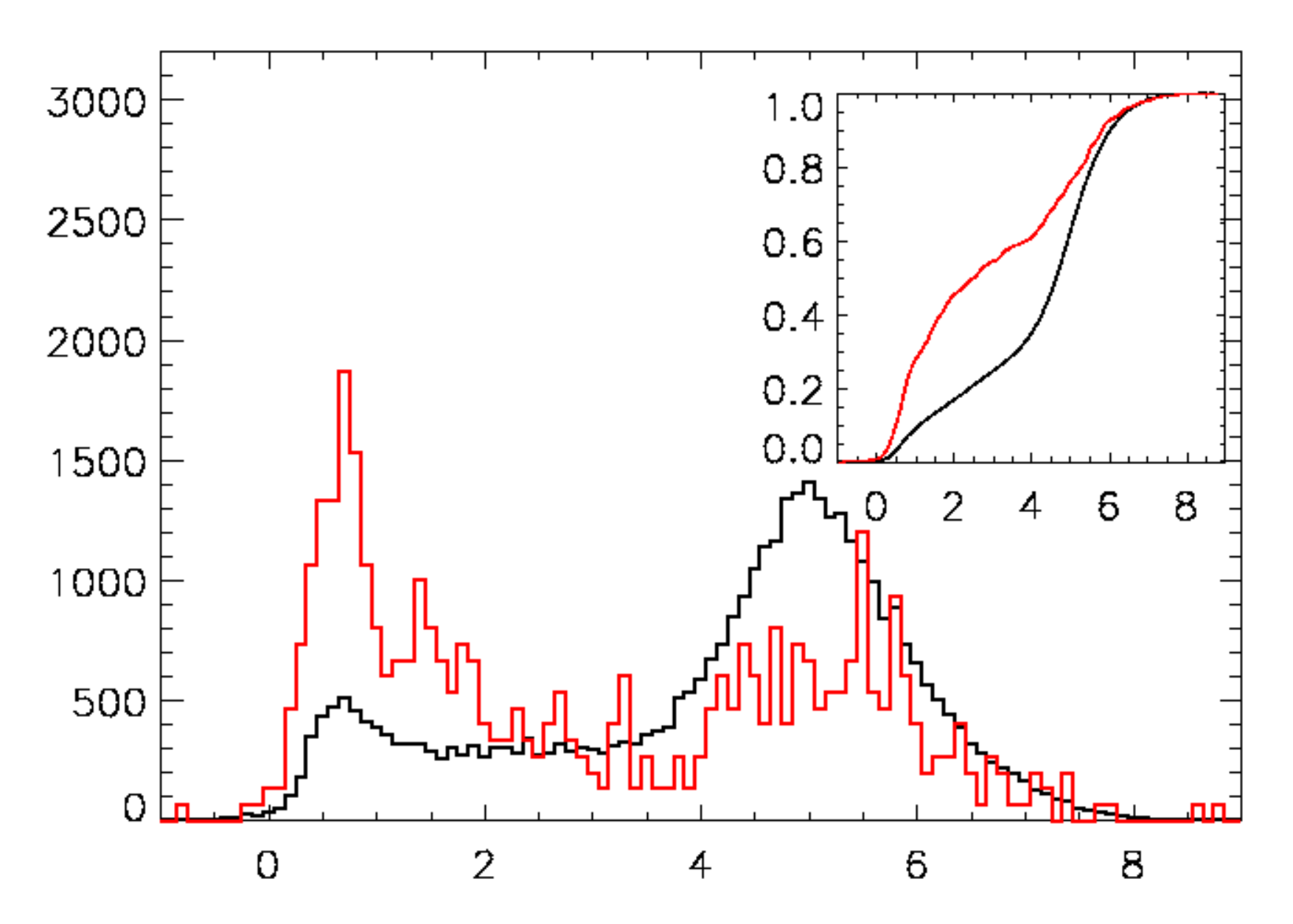}
}
\hspace{-0.6cm}
\subfigure[$(H-K_{s})$] 
{
    \label{fig:colordistributions:f}
    \includegraphics[width=5cm]{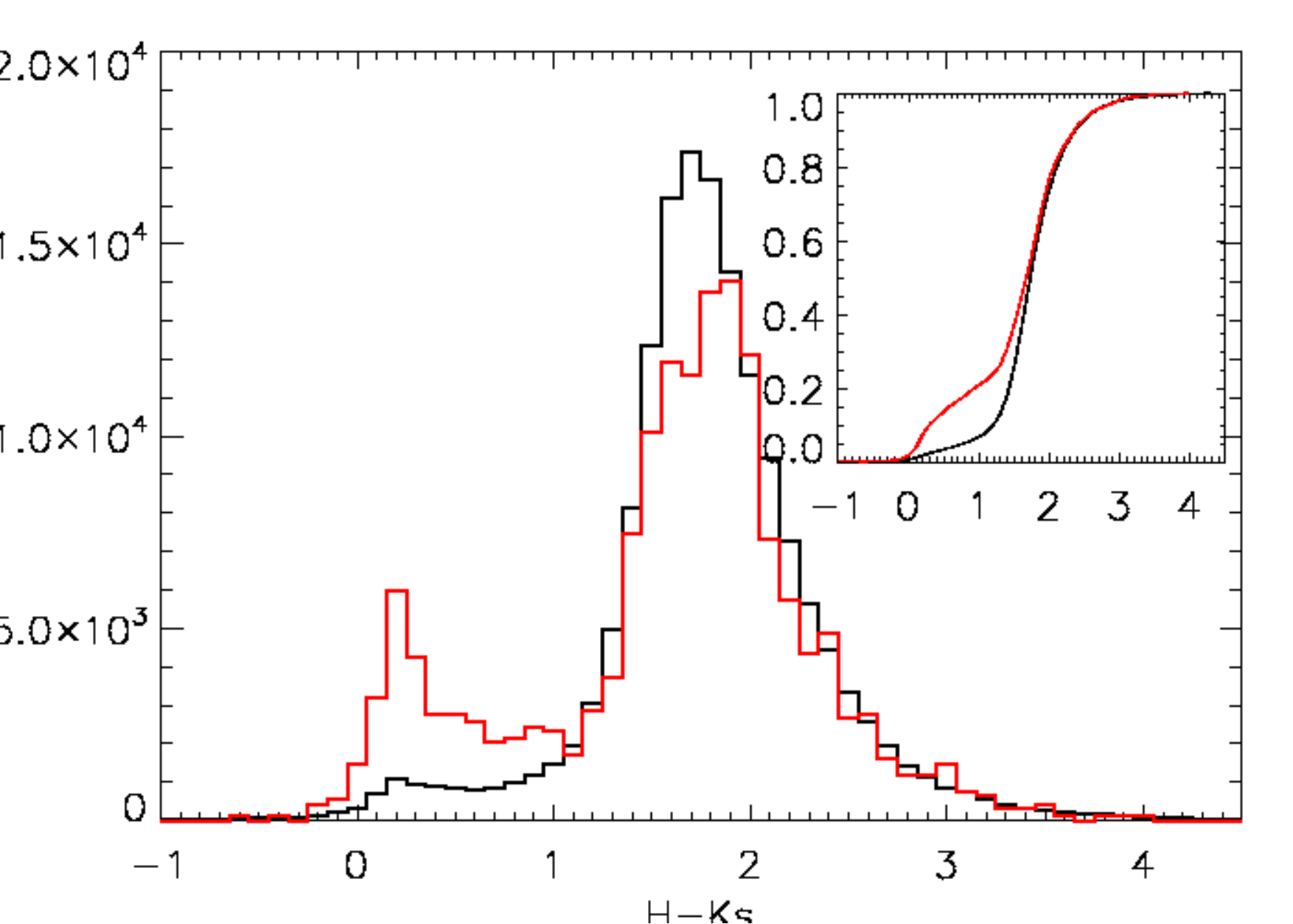}
}
\caption{Magnitude and color histograms and cumulative distributions
  for all sources in the ISPI IR catalog (in black) and for just the
  IR source matches to X-ray sources (in red). The histograms for the
  X-ray matched IR sources are normalized to the histograms of the
  entire IR catalog.}
\label{fig:maghists} 
\end{figure}

\begin{center}
\begin{figure}[ht]
\includegraphics[scale=0.75] {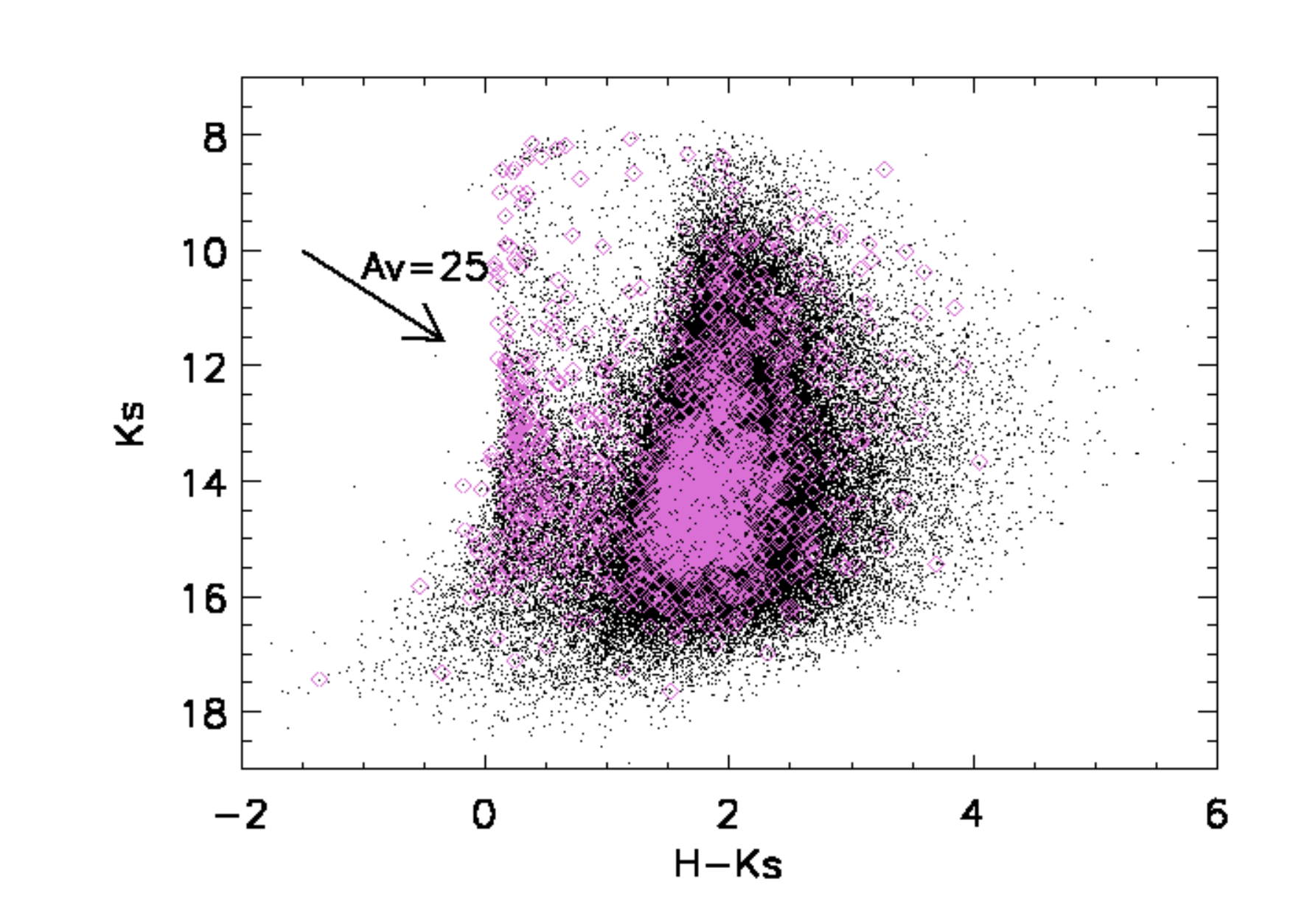}
\caption{Color magnitude diagram of the ISPI catalog (in black) with
  candidate counterparts to \textit{Chandra} X-ray sources marked in
  magenta.}
\label{fig:cmd}
\end{figure}
\end{center}

\begin{center}
\begin{figure}[ht]
\includegraphics[scale=0.75] {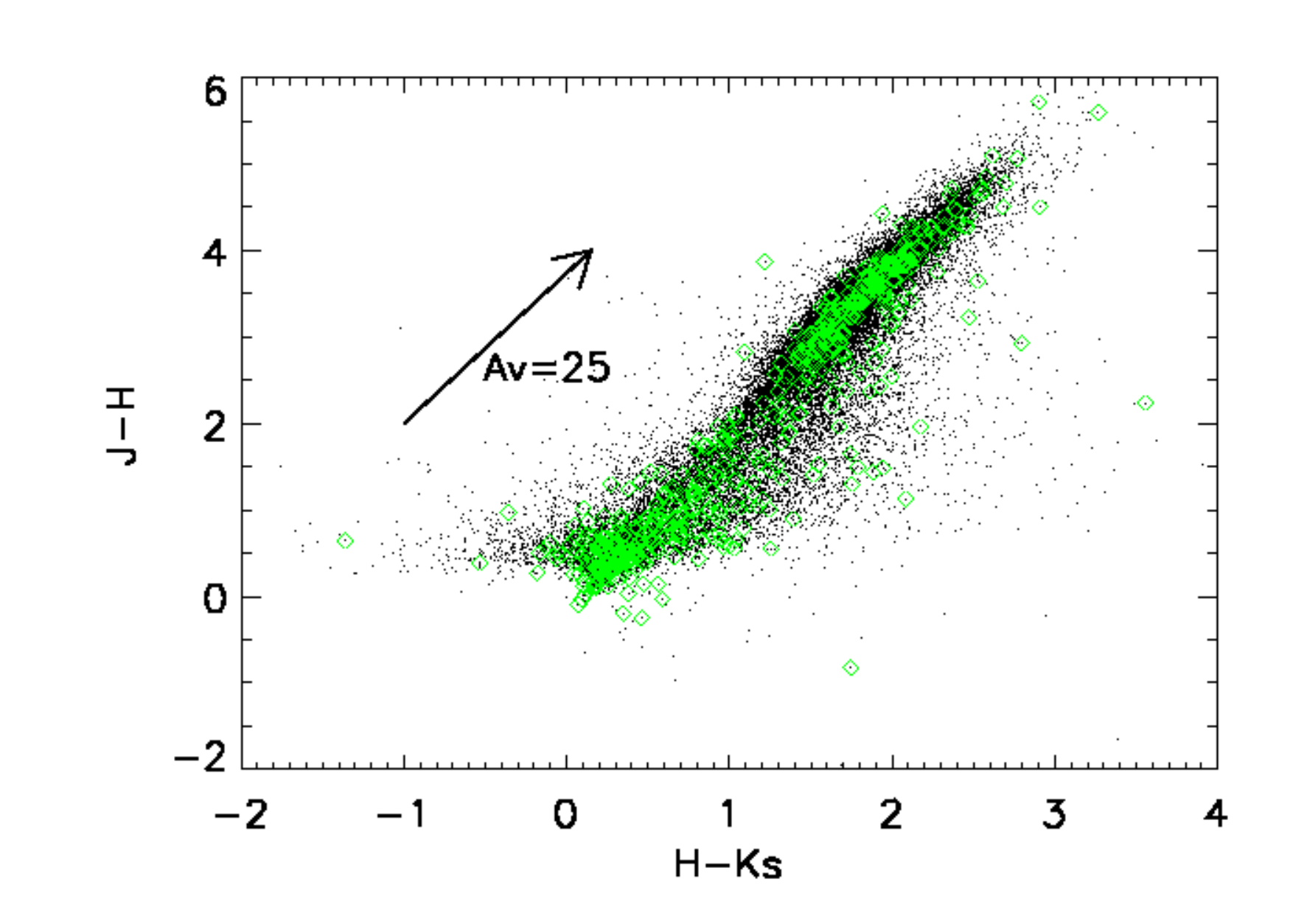}
\caption{Color-color diagram of the ISPI catalog (in black) with
  candidate counterparts to \textit{Chandra} X-ray sources marked in
  green.}
\label{fig:cc}
\end{figure}
\end{center}

\subsection{Definitions of the catalog properties}

We divide the matched catalog into properties of the X-ray and IR
sources in order to calculate the number of true physical counterparts
with a given set of IR/X-ray properties. The X-ray characteristics
include $\sigma_X$, the X-ray source positional error, in arcseconds;
the X-ray hardness ratio; the X-ray luminosity; and the number of IR
sources in the vicinity of the X-ray source.

\begin{figure}[ht]
\centering
\subfigure[] 
{
    \label{fig:errdist:left}
    \includegraphics[width=7cm]{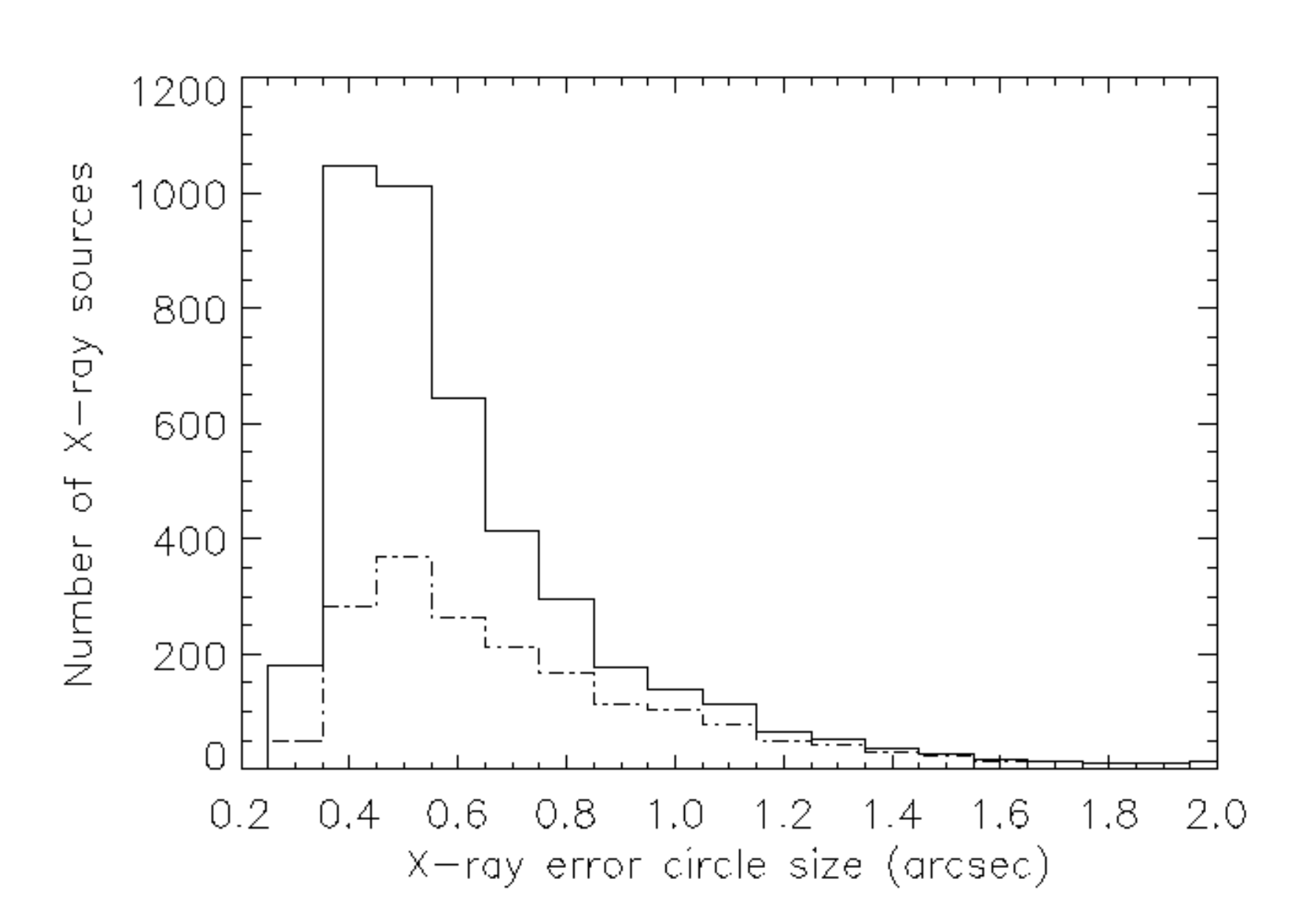}
}
\hspace{0.1cm}
\subfigure[] 
{
    \label{fig:errdist:right}
    \includegraphics[width=7cm]{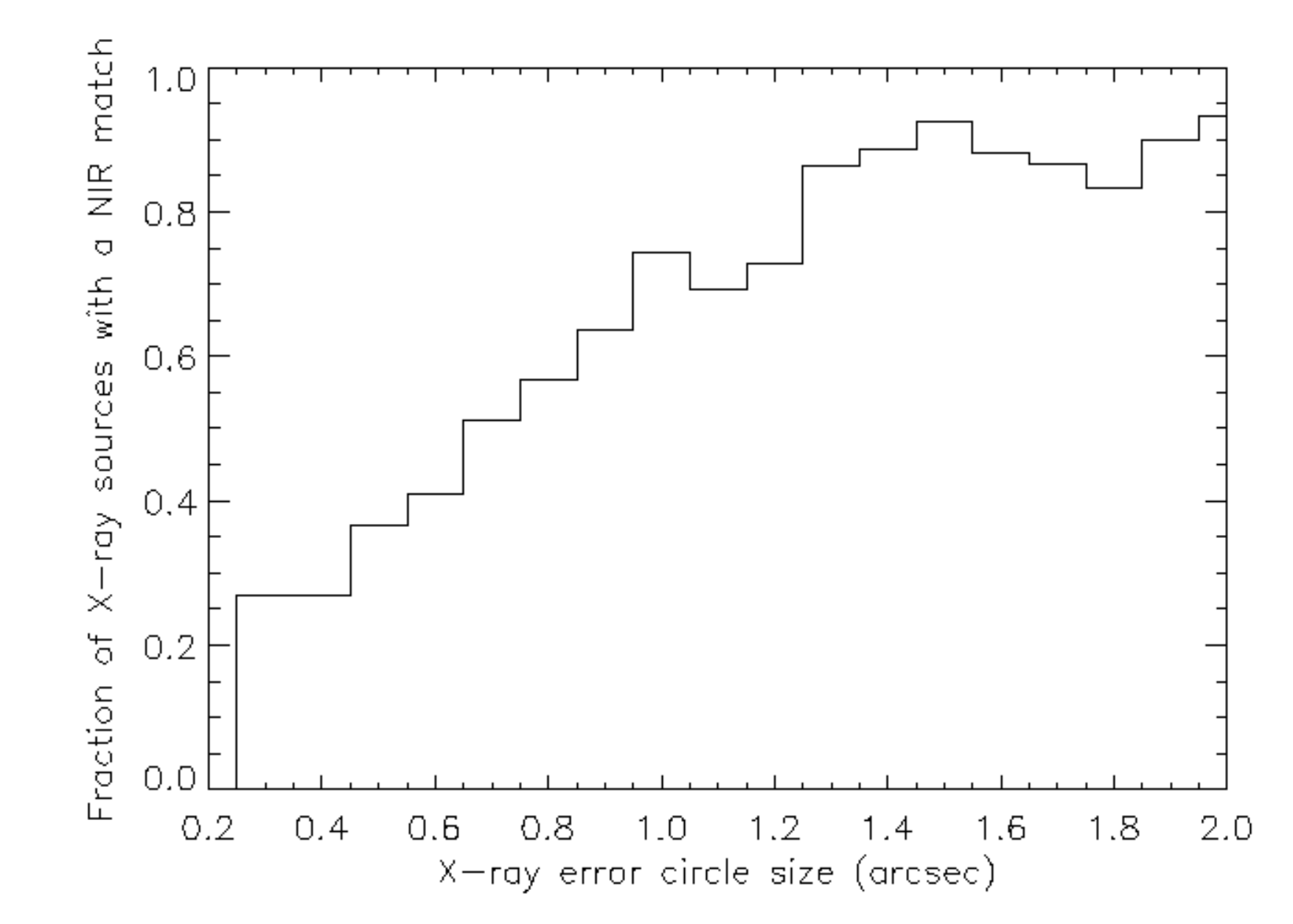}
}
\caption{(left) Distributions of positional error sizes of all 4268
  \textit{Chandra} X-ray sources in the ISPI GC field (solid line),
  and for the subset of these X-ray sources with one or more NIR
  matches (dashed line).(right) Fraction of X-ray sources with one or
  more NIR astrometric matches versus the X-ray positional error
  size.}
\label{fig:errdist} 
\end{figure}

\textit{1. Positional Error} Sources with smaller X-ray error circles
are likely to have fewer coincidental matches. In the left panel of
Figure \ref{fig:errdist} we show the histogram of position errors for
the 4268 X-ray sources with positional uncertainties under 2$"$ and
for the 1864 X-ray sources with NIR matches. The fraction of sources
with a match (shown in the right panel of Figure \ref{fig:errdist}) is
about 27 $\%$ for the X-ray sources with errors of 0.3$''$, and
reaches nearly 90$\%$ around 1.1$"$. This implies that the sources
with larger positional errors have more spurious matches, as would be
expected.

\textit{2. Hardness ratios} Soft X-rays are not expected to penetrate
the gas and dust column to the Galactic Center. The X-ray hardness
ratio provides a way to divide the X-ray sources into those that are
heavily absorbed and therefore at or beyond the Galactic Center, or
soft and therefore located well in front of the GC. We adopt the same
hardness ratio criteria as in \cite{muno09}. They define $HR0 =
\frac{(h-s)}{h+s}$, where $h$ is the flux in the 2.0-3.3 keV energy
band and $s$ is the flux in the 0.5-2.0 keV energy band. Soft sources
are defined to have $HR0 < -0.175$. Any source with a $HR0 > -0.175$
or with no counts in the soft band are defined to be \textit{hard}
sources. Of the 4268 X-ray sources in our field, 3583 are
\textit{hard} and 685 are \textit{soft}.

\textit{3. X-ray brightness} We divide the X-ray catalog into
\textit{bright} and \textit{faint} on the basis of the total count
rate. The distribution of count rates shows a roughly Gaussian
distribution with a tail toward the bright end. We define sources to
be within this bright tail if their combined hard and soft count rate
is greater than $1 \times 10^{-4}$ photons $s^{-1} cm^{-2}$.  There are
1075 sources brighter than this threshhold and 3193 fainter.  We
estimate the corresponding X-ray luminosity by assuming that the
sources lie in the Galactic Center and using the \textit{Chandra}
proposal planning tool, \textit{PIMMS}. Using a GC distance of 8kpc, a
power law index of $\Gamma=2.0$, and an absorbing column of $N_{H}=6
\times 10^{22} cm^{-2}$, our bright/faint threshold corresponds to
$L_{X}=8 \times 10^{31}$ ergs/s for sources at the GC distance. 

\textit{4. IR source density} Dust lanes are evident in the image of
the ISPI data in Figure \ref{fig:colorimage}. The lanes correspond to
lines of sight with higher extinction and are seen as regions with
more reddened colors and fewer star counts.  We attempt to
quantitatively locate these higher extinction areas by computing the
average stellar surface density in the vicinity of the X-ray
sources. We calculate the stellar surface density using stars with $10
<~K_{s}~<~13.5~mag$ within a 10" radius of each X-ray source and
normalize this value by $\Sigma_{ISPI}$, the mean density of
$10<~K_{s}~<~13.5~mag$ stars in the entire ISPI GC field. Therefore
the density parameter is the fraction of the mean density of the
field.  We call regions with $> 1.0\times\Sigma_{ISPI}$,
\textit{dense} and regions with values $\le~ 1.0 \times
\Sigma_{ISPI}$, \textit{sparse}. The intention is that \textit{sparse}
areas indicate regions of high extinction and \textit{dense} areas
indicate windows of lower extinction.

\begin{figure}[ht]
\centering
\subfigure[$(J-H)$\textit{ all matches}] 
{
    \label{fig:colordistributions:a}
    \includegraphics[width=5cm]{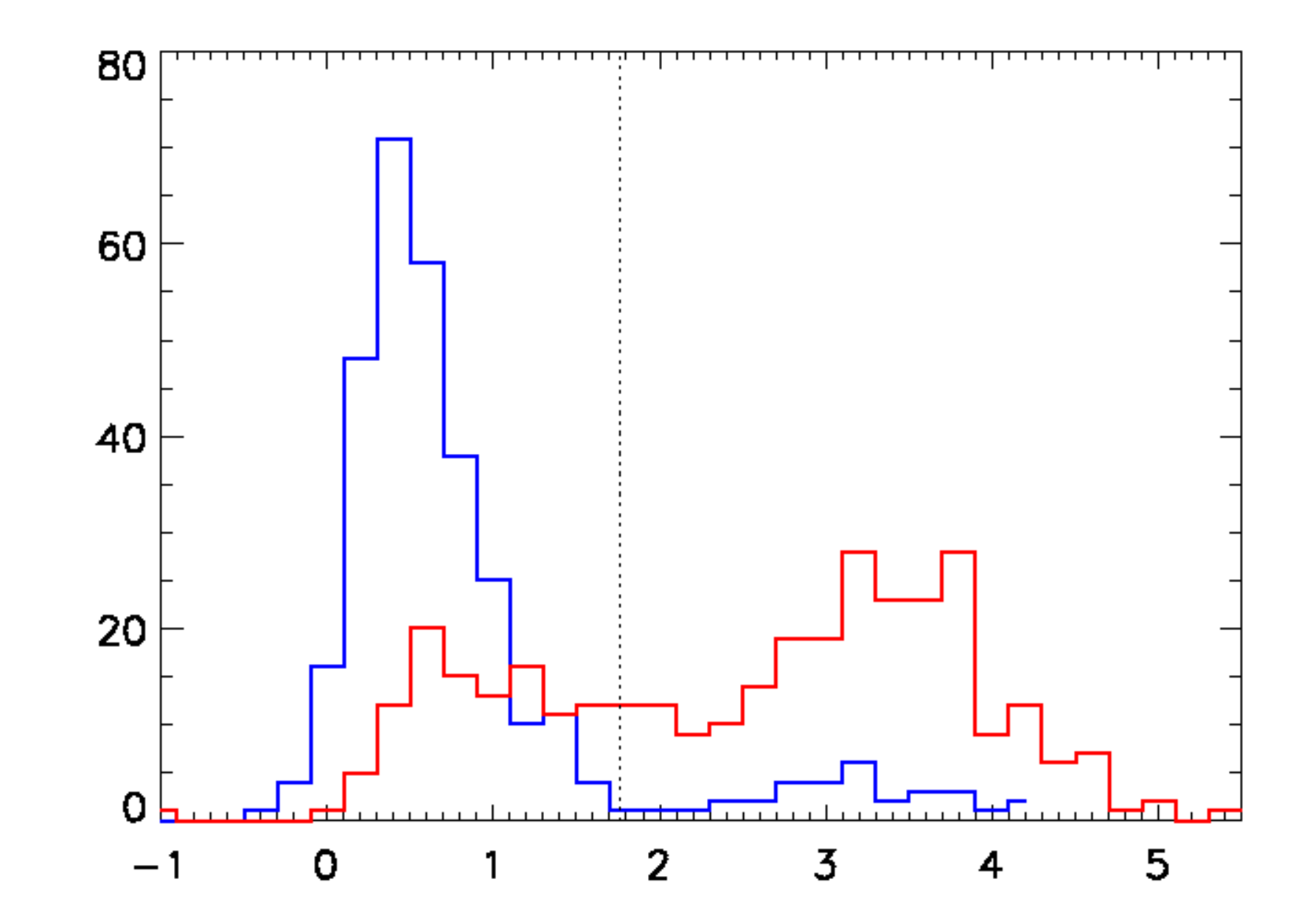}
}
\hspace{-0.8cm}
\subfigure[$(J-K_{s})$\textit{ all matches}] 
{
    \label{fig:colordistributions:b}
    \includegraphics[width=5cm]{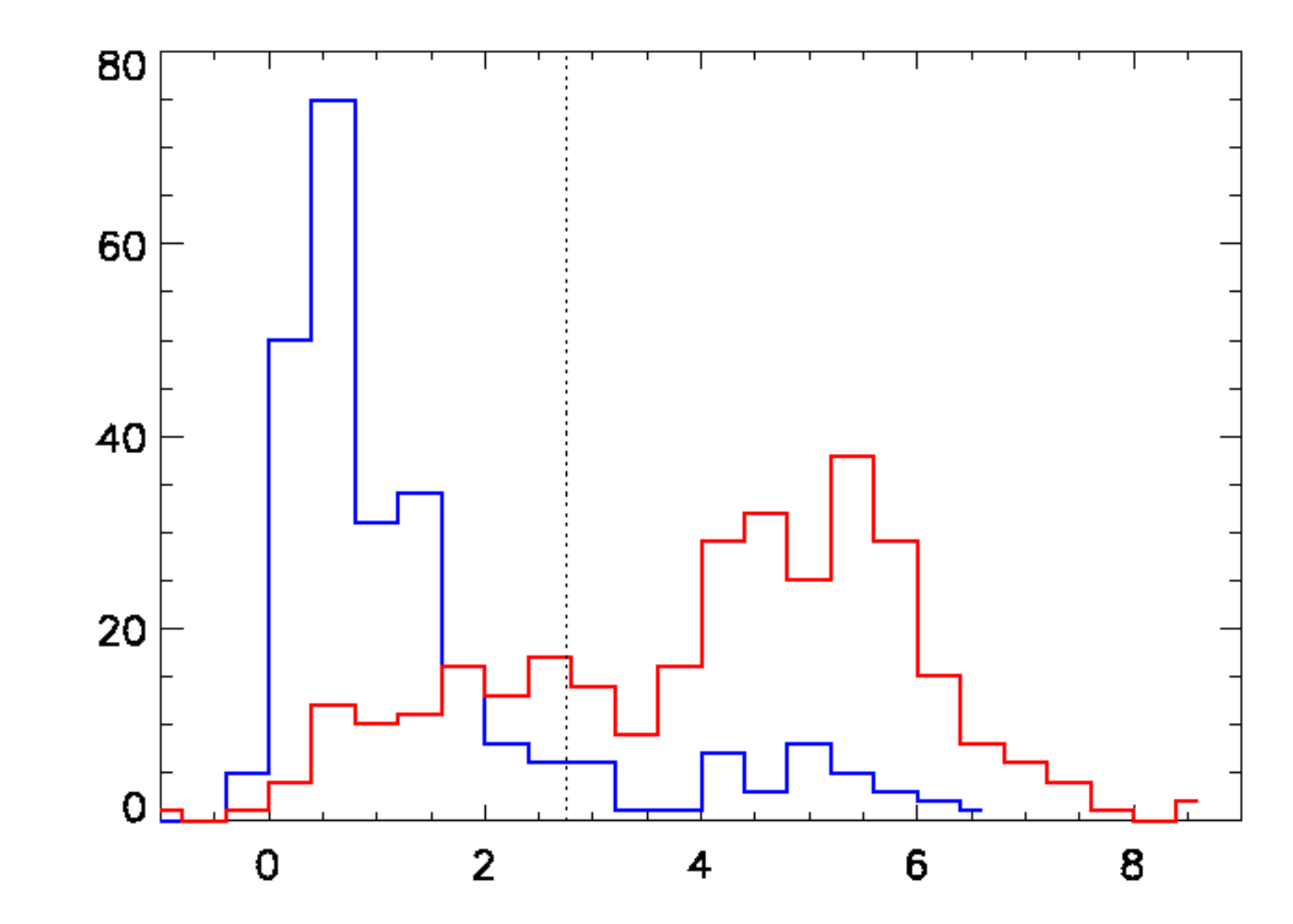}
}
\hspace{-0.8cm}
\subfigure[$(H-K_{s})$\textit{ all matches}] 
{
    \label{fig:colordistributions:c}
    \includegraphics[width=5cm]{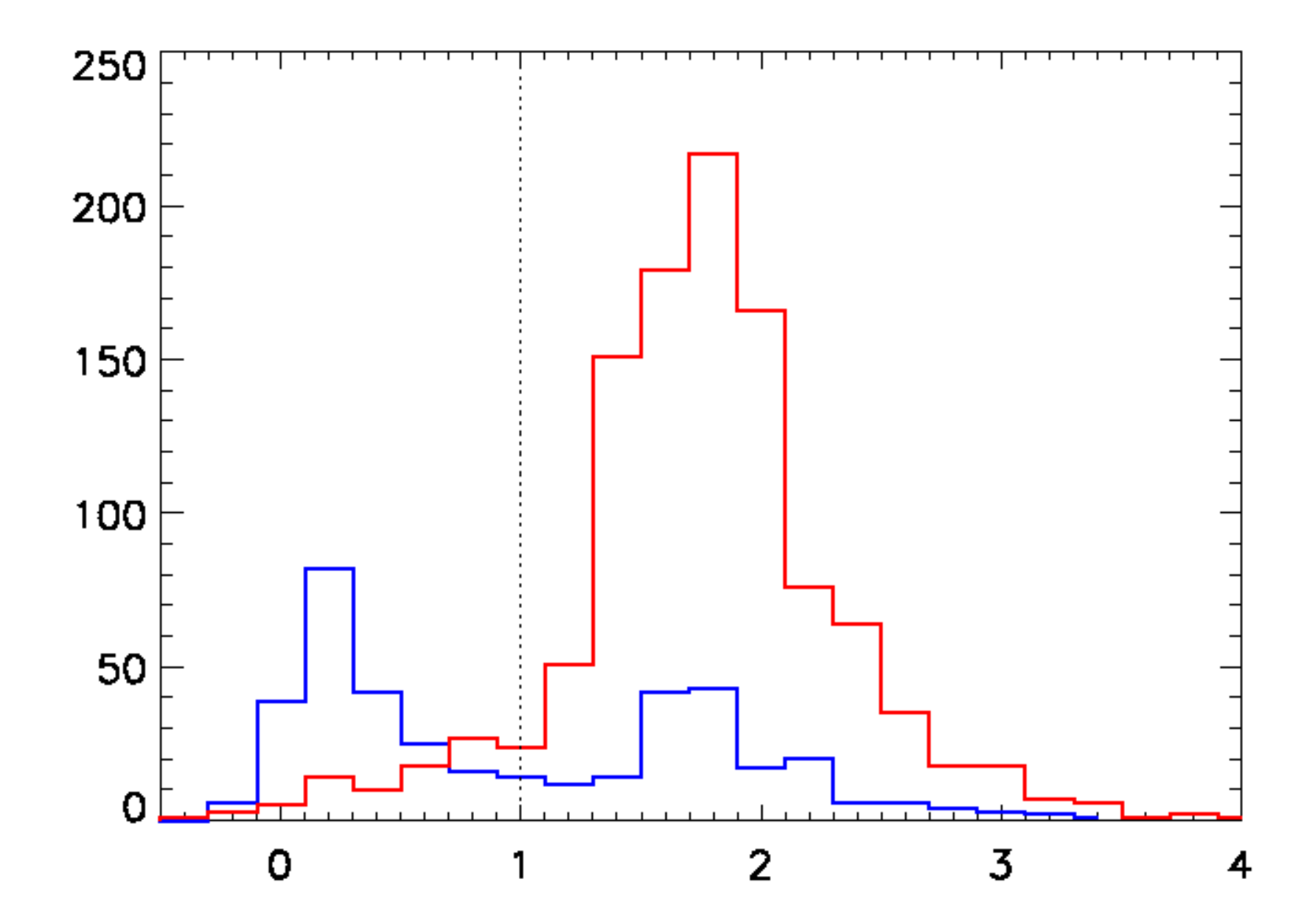}
}
\hspace{-0.8cm}
\subfigure[$(J-H)$\textit{ real matches}] 
{
    \label{fig:colordistributions:d}
    \includegraphics[width=5cm]{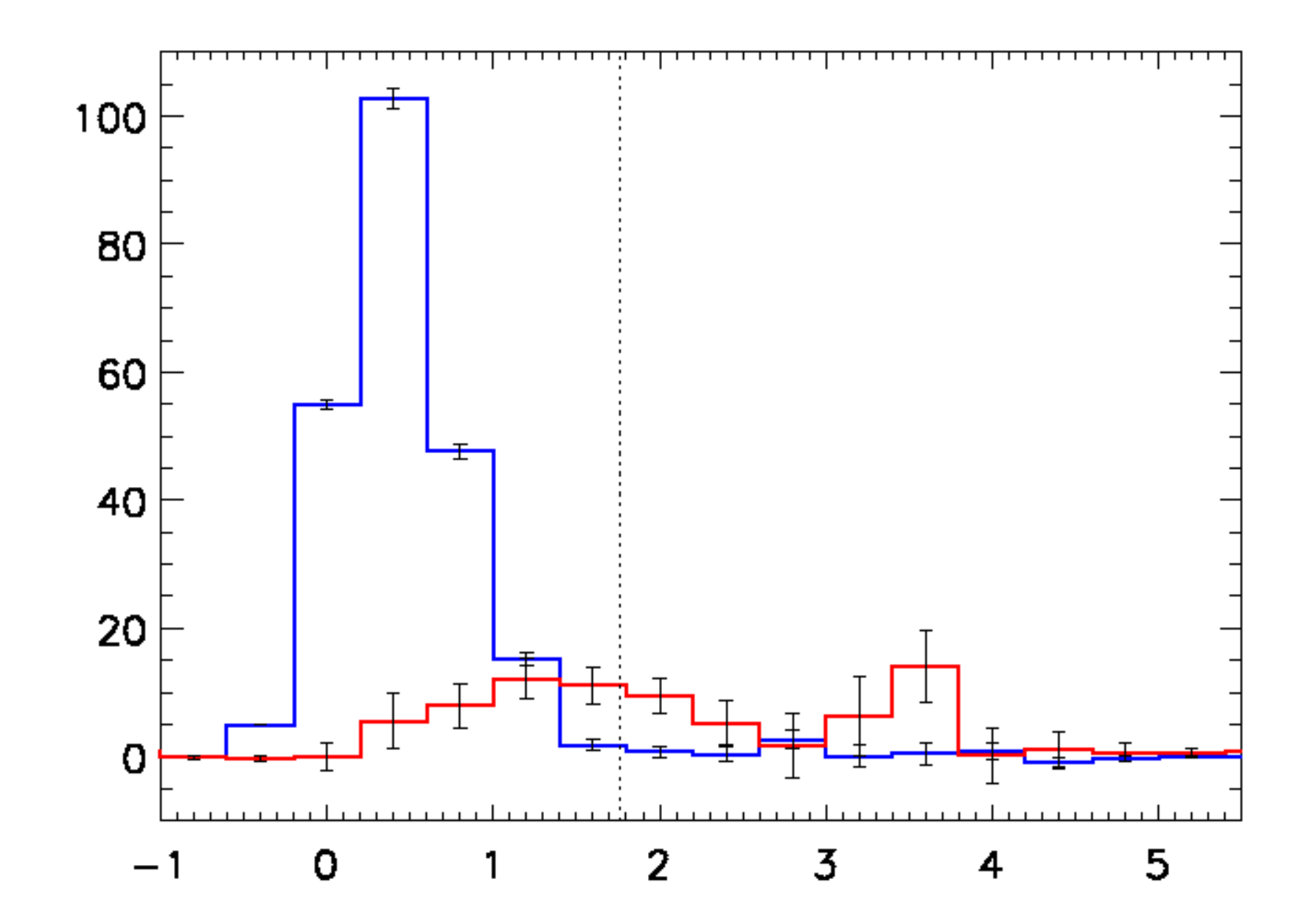}
}
\hspace{-0.8cm}
\subfigure[$(J-K_{s})$\textit{ real matches}] 
{
    \label{fig:colordistributions:e}
    \includegraphics[width=5cm]{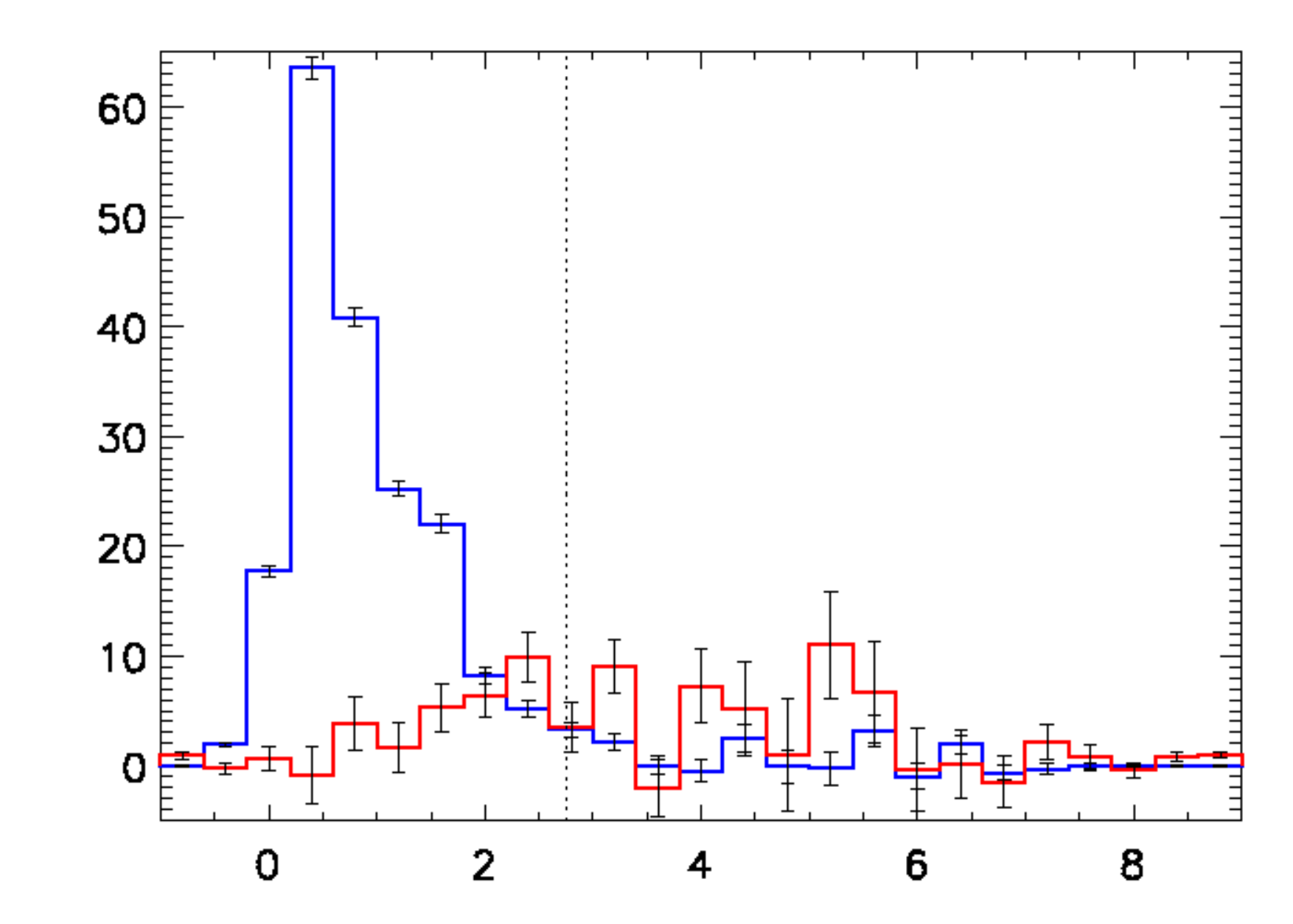}
}
\hspace{-0.8cm}
\subfigure[$(H-K_{s})$\textit{ read matches}] 
{
    \label{fig:colordistributions:f}
    \includegraphics[width=5cm]{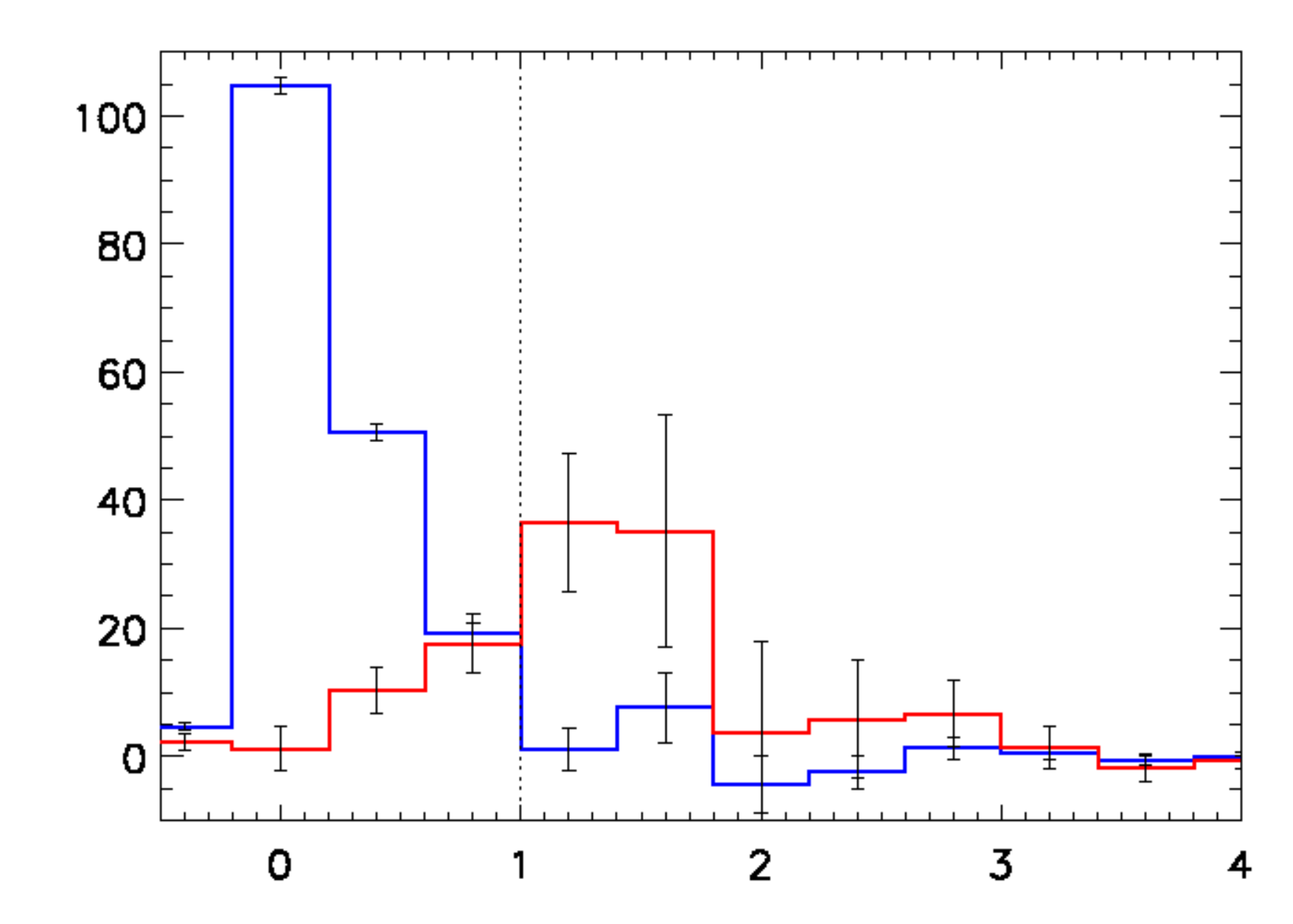}
}
\caption{Histograms of the infrared colors of matches to soft (blue
  histogram) and hard (red histogram) X-ray sources. Frames (a), (b)
  and (c) show the $J-H$,$J-K_{s}$ and $H-K_{s}$ distributions for
  \textit{all} candidate counterparts and frames (d),(e) and (f) show
  the same color distributions for the estimated numbers of real
  counterparts (see Sections \S4.5 and \S4.6). The dotted lines
  indicate our cut between \textit{unreddened} and \textit{reddened}
  ISPI sources.}
\label{fig:colordistributions} 
\end{figure}

\textit{5. IR catalog properties} The characteristics of the infrared
catalog include the $J$, $H$ and $K_{s}$ magnitudes and the $J-H$,
$J-K_{s}$, and $H-K_{s}$ colors.  Two major populations can be seen in
the $H-K_{s}$ distribution in Figure
\ref{fig:colordistributions}. There is a less reddened population,
centered at $H-K_{s}=0.2~mag$ and a more reddened population centered
at $H-K_{s}=2.0~mag$. We separate the two at $H-K_{s}=1.0~mag$, and
call sources with $H-K_{s} < 1.0~mag$, \textit{unreddened} and sources
with $H-K_{s} > 1.0~mag$, \textit{reddened}. We derived corresponding
criteria for $J-H$ and $J-K_{s}$ using the Galactic bulge extinction
ratios from \cite{nishiyama08}. These were $J-H < 1.76~mag$, $J-K_{s}
< 2.76~mag$ for \textit{unreddened} sources and $J-H > 1.76~mag$,
$J-K_{s} > 2.76~mag$ for \textit{reddened} sources.

\subsection{Estimating the number of spurious NIR matches to the \textit{Chandra} catalog}

We simulated the rate of false matches by performing Monte Carlo tests
of the matching procedure. In this procedure, the positions of the
ISPI NIR catalog sources are fixed while the positions of the 4268
X-ray sources with $\sigma_{X}\leq 2.0"$ are randomly shifted. The
X-ray sources are repositioned to lie between $0"$ and $10"$ of their
original position in a random direction. All the other X-ray
properties, including the positional error and X-ray photometry are
preserved during the process. We performed the randomization 3,000
times, producing 3,000 artificial X-ray catalogs. Then we
cross-correlated the artificial X-ray catalogs to the NIR catalog in
the same way as for the original, positionally aligned X-ray catalog,
using $\sigma_{X} + 0.2"$ as the matching radius, in arcseconds. In
the following, we call the X-ray catalog with the true X-ray source
positions the \textit{aligned} X-ray catalog.

The number of NIR-matched X-ray sources to the aligned X-ray catalog
is called $N_{obs}$ and the total number of X-ray sources in the
catalog is $A$.  The \textit{average} number of IR-matched X-ray
sources to the randomized catalogs is called $N_{ran}$, and the
standard deviation of this number is called $\sigma_{N_{ran}}$. These
variables can refer to the entire catalog of X-ray sources or to a
subset with specified X-ray properties (for example, \textit{soft} X-ray
sources with \textit{bright} X-ray flux, or \textit{hard} X-ray
sources with $\sigma_{X}\leq 0.5"$).

The number $N_{obs}$ includes both real physical matches and spurious
matches. We would like to calculate the number of spurious matches,
$N_{spur}$ and subtract it from $N_{obs}$ to obtain the number of real
matches, $N_{real}$. However, the randomized X-ray catalogs consist of
randomizations of \textit{all} the X-ray sources, including the ones
with true detected NIR counterparts. When randomized, these true
matches contribute to the total $N_{ran}$, making it an
\textit{overestimate} of the number of spurious matches to the aligned
X-ray catalog.

Fortunately, $N_{real}$ can be solved in terms of the known values of
$N_{obs}$, $N_{ran}$ and $A$. The variable definitions are collected
in Table \ref{tab:vardefs1}.

From the definition of $N_{obs}$: 
\begin{equation}
N_{obs}= N_{real}+N_{spur}
\end{equation}

To represent $N_{ran}$ in terms of our defined variables we assume
that the randomized X-ray catalogs should contain 0 real NIR matches
to the X-ray sources. We define the rate of spurious IR matches to the
real X-ray catalog to be $R$. This rate is the number of spurious
matches divided by the total number of X-ray sources \textit{without}
detected real NIR counterparts, i.e. $A-N_{real}$.
\begin{equation}
R= \frac{N_{spur}}{A-N_{real}}
\end{equation}
If we assume that $R$ is the same value for \textit{all} the
randomized X-ray catalogs, then $N_{ran}$ is simply, $A \times R$.
\begin{equation}
N_{ran} = A \times \frac{N_{spur}}{A-N_{real}}
\end{equation}
When equations 1 and 3 are solved, we find that the number of real NIR matches within $N_{obs}$ is:
\begin{equation}
N_{real}= N_{obs} - \frac{N_{ran}(A-N_{obs})}{A-N_{ran}}
\end{equation}
and the number of spurious matches is 
\begin{equation}
N_{spur}= \frac{N_{ran}(A-N_{obs})}{A-N_{ran}}
\end{equation}

The uncertainty in these numbers is calculated by standard error
propagation using the standard deviation of the number of matches from
the random simulations, $\sigma_{N_{ran}}$.
\begin{equation}
\sigma_{N_{real}}=\sigma_{N_{spur}}=\frac{A^{2}-AN_{obs}}{(A-N_{ran})^{2}}\sigma_{N_{ran}}
\end{equation}

\begin{table}[ht]
\begin{center}
\caption{Variable definitions for calculating numbers of real counterparts to X-ray sources, with unspecified IR properties}
\label{tab:vardefs1}
\small
\begin{tabular}{crrrrrrrrrrr}
  \tableline\tableline
  Variable name & Meaning\\
  \tableline
  $A$              &  total num. of X-ray sources with a given set of X-ray properties  \\
  $N_{obs}$        &  num. of singly matched X-ray sources from aligned X-ray catalog\\ 
  $N_{ran}$        & mean num. of singly matched X-ray sources from rand. X-ray catalogs\\
  $\sigma_{N_{ran}}$      & standard deviation of $N_{ran}$ \\
  $N_{real}$         & num. of \textit{real} IR counterparts within $N_{obs}$ matches \\
  $N_{spur}$         & num. of \textit{spurious} IR counterparts within $N_{obs}$ matches \\
  $\sigma_{N_{real}}$& $1\sigma$ error in value of $N_{real}$\\
  $\sigma_{N_{spur}}$& $1\sigma$ error in value of $N_{spur}$\\
  \tableline
\end{tabular}
\end{center}
\end{table}

We note that there are two critical assumptions for equations 1-6.  

1) $N_{obs}$ and $N_{ran}$ contain only X-ray sources matched to a
single IR counterpart. Our prescription omits the $\sim 6\%$ of X-ray
sources with multiple matches, which obviously can also contain real
X-ray/NIR counterparts. We address this omission in \S 5, where we use
the rate of matches to real counterparts from these equations to
estimate the number of additional real NIR counterparts present in the
X-ray sources matched to multiple IR sources.

2) Equations 1-6 can only be used to find the numbers of real IR
matches to X-ray sources \textit{with specified X-ray properties}. The
situation becomes more complicated if we wish to find the number of
X-ray sources that have real IR counterparts \textit{with a specified
  IR property}. In these cases, the properties of the real NIR
counterparts can bias the matching simulations. For example, we found
that the number of soft X-ray sources matched to reddened IR
counterparts was \textit{higher} for the simulations than for the real
X-ray catalog. This is because there is a number of real unreddened
NIR counterparts to the soft X-ray sources, and when they are
positionally randomized, they are more likely to be matched to
reddened NIR sources, which make up $\sim$80$\%$ of the ISPI
catalog. The solution is to rederive equations 1-6, keeping track of
both the number of sources with and without the specified NIR
property. This derivation is presented in \textit{Appendix A}.

In the next section we briefly ignore the biases associated with the
matching simulations in order to get an illustrative view of the
content of the matched X-ray/NIR catalog. Then we apply these
corrective equations in \S 4.5-4.6 and \S 6 to find more exact numbers of the
real IR/X-ray counterparts.

\subsection{The fractional excess of X-ray sources with NIR matches}

We compare the fraction of matched X-ray sources in the Chandra
catalog to the fraction obtained with the randomized catalogs. The
randomized catalogs should contain essentially no true matches, so the
difference between these two fractions will yield the fraction of
X-ray sources with authentic detected matches in the ISPI
catalog. Since this calculation does not address the overcounting of
randomized catalog matches described in \S 4.2, the rates are all likely
to be underestimates but they are useful to get an illustrative view
of the level of real and spurious matching between the Chandra and
ISPI catalogs. 

We sort the X-ray catalog by the properties described in \S 4.1,
including hard/soft, bright/faint and dense/sparse, and we count the
number of sources with a specific set of properties that are matched
to NIR sources. The size of the X-ray positional error circle is a
strong factor in whether an X-ray source is matched, so we record the
number of matched and unmatched sources per each X-ray positional
error bin ($\sigma_{X}$ is between $0.3"$ and $2.0"$ in intervals of
$0.1"$). We call the number of matched sources per error bin
$N_{obs,i}$ where $i$ is a label for the X-ray error circle size. The
mean number of matched sources to the randomized X-ray catalogs is
$N_{ran,i}$ and its standard deviation is $\sigma_{N}$. We divide
$N_{obs,i}$,$N_{ran,i}$, and $\sigma_{N}$ by $A_{i}$, the total number of
X-ray sources in the Chandra catalog with the specified properties and
error circle size. This yields the matching fractions $f_{obs,i}$ and
$f_{ran,i}$ and the $1\sigma$ fractional scatter, $\sigma_{f,i}$. The
fraction of X-ray sources (with a given $\sigma_{X}$) with a real
detected counterpart is:
\begin{equation}
  f_{real,i}=f_{obs,i}-f_{ran,i}
\end{equation}

We combine the values of $f_{real,i}$ by calculating the weighted mean
using $\sigma_{f,i}$ as the weight.

\begin{equation}
  f_{real}=\frac{\sum\limits_{i}\frac{f_{obs,i}-f_{ran,i}}{\sigma_{f,i}^{2}}}{\sum\limits_{i}\frac{1}{\sigma_{f,i}^{2}}}
\end{equation}

$i$ is a label for the X-ray positional error bins (e.g. $i=1,2,3$...$\Rightarrow$ $\sigma_{X}=0.3",0.4",0.5"$...)

The formal uncertainty of this expression is:
\begin{equation}
  \sigma_{f}=\sqrt{\frac{1}{\sum\limits_{i}\frac{1}{\sigma_{f,i}^{2}}}}
\end{equation}

\begin{table}[ht]
\begin{center}
\caption{Relative matching fraction for X-ray sources}
\label{tab:matchfraction}
\begin{tabular}{crrrrrrrrrrr}
\tableline\tableline

Criteria & $f_{real} \pm \sigma_{f}$ & Significance \\
\tableline
All X-ray sources            & 0.052 $\pm$ 0.007  & 7.7$\sigma$ \\ 
Soft sources                 & 0.22 $\pm$ 0.02    & 13.6$\sigma$ \\
Hard sources                 & 0.016 $\pm$ 0.007  & 2.2$\sigma$ \\
Soft, faint sources          & 0.15 $\pm$ 0.02    & 8.0$\sigma$ \\
Soft, bright sources         & 0.42 $\pm$ 0.03    & 13.3$\sigma$ \\
Soft sources, dense regions  & 0.18 $\pm$ 0.02    & 8.3 $\sigma$ \\
Soft sources, sparse regions & 0.23 $\pm$ 0.02    & 10.4 $\sigma$ \\
Hard, faint sources          & 0.001$\pm$ 0.008   & 0.2$\sigma$ \\
Hard, bright sources         & 0.05 $\pm$ 0.01    & 3.8$\sigma$ \\
Hard sources, dense regions  & 0.012$\pm$ 0.009   & 1.4 $\sigma$ \\
Hard sources, sparse regions & 0.02 $\pm$ 0.01    & 1.7 $\sigma$ \\
\tableline
\end{tabular}
\end{center}
\end{table}

In Table \ref{tab:matchfraction} we show the $R_{real}$ values and
uncertainties for combinations of the X-ray source properties. 22$\%$
of soft sources have detected matches in the ISPI catalog compared to
only 1.6$\%$ of the hard sources. Both soft and hard sources show a
bias toward detection of \textit{bright} sources over \textit{faint}
sources, with a \textit{bright} source being more than twice as likely
to have a detected counterpart. 

There is a slight ($6\% \pm 3\%$) bias toward detecting soft sources
in \textit{sparse} regions over \textit{dense} regions. This effect
may be caused by the crowding limits being more severe in
\textit{dense} regions. There is no significant difference between
\textit{dense} and \textit{sparse} regions seen in the hard
sources. However, the low number statistics for hard sources would
mask an effect of similar magnitude to the soft sources.

The main point to draw from this is that hard sources have a
particularly low rate of IR detection in the data, at least 4 times
lower than for soft sources. We conclude that most of the sources in
the X-ray catalog are too faint in the NIR to be detected by the ISPI
observations, which is consistent with the finding that 57$\%$ of
X-ray sources do not have a candidate match at all. We discuss the
charactistics of the sources without NIR matches in the next section.

\subsection{Characteristics of the X-ray sources without NIR matches}

57 $\%$ of the X-ray catalog sources have no astrometric matches in
the ISPI catalog. Soft X-ray sources have an unmatched percentage of
31$\%$ (214 unmatched out of 685 total soft X-ray sources) and hard
X-ray sources have an unmatched percentage of 62$\%$ (2211 unmatched
out of 3583 total hard X-ray sources).

The unmatched sources are skewed toward having smaller error circles
compared to the total X-ray catalog. Their mean error circle size is
$0.55"$ compared to $0.77"$ for the matched sources. We expect sources with
smaller error circles to be less likely to have chance astrometric
matches.

We find no difference in the X-ray brightness distributions of the
unmatched X-ray sources and total catalog. However, we do find that
the mean IR density in the vicinity of unmatched sources is
\textit{higher} than for the entire catalog. The mean stellar surface
density around unmatched X-ray sources is $1.6~\times~\Sigma_{ISPI}$,
compared to $1.46~\times~\Sigma_{ISPI}$ for the entire X-ray
catalog. The effect is probably due to the fact that the smallest
error circles with the lowest probability of chance matches lie in the
center of the field where the stellar density is highest. For sources
with $\sigma_{X} \leq 0.5"$ and $\sigma_{X} > 0.5"$, the mean
densities are $1.8 \times \Sigma_{ISPI}$ and $1.0 \times
\Sigma_{ISPI}$, respectively. Stellar density should have a linear
effect on the chance for a spurious match, but the radius of the X-ray
error circle should have a quadratic effect. Thus, it is predictable
that the sources in the densest areas are also the most frequently
unmatched.

We demonstrated in \S 4.3 that most of the astrometrically matched
X-ray sources are spurious matches which means that either the source
is too faint to be detected at all or it is blended with the brighter
source or sources to which they are coincidentally matched. The
unmatched sources can be useful because we know that in most cases
their magnitudes are fainter than the ISPI data's detection limit. The
ISPI detection limit varies with crowding conditions but for this
discussion we adopt the mean $50 \%$ completeness limits from
artificial star tests in Table \ref{tab:catalogcharacteristics}, which
are $J=18.4~mag$, $H=16.4~mag$ and $K_{s}=14.5~mag$.

The unmatched soft sources are likely to be coronally active late type
stars \citep{muno03a}. Coronal activity at the $>10^{27}$ ergs$/$s
level is seen at all spectral types in field dwarf stars
\citep{feigelson04}. For an illustrative case, we consider that a
$J=9~mag$ M0 V star with no extinction would be detected in our ISPI
data out to a distance of $\sim$ 800pc. \cite{muno03a} determine that
the soft sources in the \textit{Chandra} GC data can reside as far out as
4kpc, which leaves a substantial volume for coronally active late type
dwarf stars to reside and yet be undetectable in our ISPI imaging.

The unmatched hard X-ray sources have a number of diverse candidates:
AGNs, isolated pulsars and magnetars, magnetic CVs and LMXBs.
\cite{muno03a} calculate that the $17' \times 17'$ central GC field
should have between 20 and 100 AGN contributing to the hard source
counts. AGN that are seen through the GC will be very
faint. \cite{bandyopadhyay05} note that there are no AGN in the
\textit{Hubble} Deep Field North survey brighter than
$K=17~mag$. After $A_{K} > 2.5~mag$ extinction is added on, this
magnitude is well beyond our sensitivity, which means that there
should be no detected counterparts for the $\sim 10^{2}$ AGN in the
X-ray catalog.
 
Isolated neutron stars can be bright at very young ages, but except
for the Crab Pulsar, the brightest $K$ band detected neutron star is
the magnetar 1E 1547.0-54.08 with $K=18.2~mag$, at a distance of 9 kpc
\citep{mignani09}. We conclude that for all but the most exceptional
systems, isolated pulsars in the GC should be undetectable in our ISPI
data.

LMXBs and magnetic CVs share the same types of donor stars, usually
late type K dwarfs \citep{jonker04,knigge06}. When in quiescence,
these systems can be dominated by the light of the secondary star. In
the case of a K0 dwarf star, a system in the Galactic Center would be
$K\sim 21~mag$, assuming $A_{K}=2.5~mag$ \citep{baganoff03} and a GC
distance of 8kpc \citep{gillessen09,cox00,carroll96}. Thus, the
majority of canonical LMXBs and CVs would be undetected at GC
distances in our ISPI data.

\subsection{Probable matches to soft X-ray sources}

The fraction of soft sources that are matched in section \S 4.3,
suggests that a relatively high number of soft source candidate
counterparts are real. We use the results of the random matching
simulations with Equations 4 and A3 to derive the numbers of real
counterparts to soft sources. We break down the soft X-ray/IR matches
by source properties in Table \ref{tab:softmatch}.

There are 387 single IR matches to X-ray soft sources. Analysis of the
randomized X-ray catalog matching suggests that $230 \pm 12$ are
probable real matches, for a rate of $60 \pm 3 \%$. We see that
almost all ($96\%$) the soft X-ray sources matched to unreddened NIR
sources with a $J$ detection are real, while a trivial number of
matches to reddened X-ray sources are real. The lack of true
soft/reddened sources is not surprising, because the X-ray attenuation
implied by the reddened IR colors should make an intrinsically soft
source be detected as a hard source, or be completely extinguished.

\begin{table}[ht]
\begin{center}
\caption{Matching statistics for soft X-ray sources matched to a single IR source}
\label{tab:softmatch}
\begin{tabular}{crrrrrrrrrrr}
\tableline\tableline
Criteria & $N_{obs}$ & $N_{ran}$ & $N_{real}$ & $R = \frac{N_{real}}{N_{obs}}$& $\sigma_{R}$& $\frac{R}{\sigma_{R}}$\\
\tableline
All soft X-ray       &  387 & 236.0  &  230  & 0.60 & 0.03 & 18.5 \\
Unreddened           &  254 & 45.9   &  224  & 0.88 & 0.02 & 47.0 \\
Reddened             &  133 & 190.1  &  7    & 0.05 & 0.08 & 0.6  \\
With J detection     &  286 & 47.8   &  254  & 0.89 & 0.02 & 51.2 \\
With no J detect.    &  101 & 188.1  & -24   &-0.24 & 0.11 & 2.2  \\   
Unred. with J det.   &  242 & 15.5   &  232  & 0.96 & 0.01 & 87.1 \\
Red. with J det.     &  44  & 32.4   &  22   & 0.51 & 0.09 & 5.6  \\
\tableline
\end{tabular}
\end{center}
\end{table}

\subsection{Probable matches to hard X-ray Sources}

There are 1223 hard X-ray sources with only a single NIR candidate
counterpart in our ISPI catalog. However, when the randomized X-ray
catalogs are matched to the ISPI catalog, 1135.0 matches are created
on average, meaning a large percentage of hard, matched sources are
spurious. Our analyses indicate that the number of real physical
associations within the 1223 matches is 129 $\pm$ 39 sources, or
$11 \% \pm 3 \%$ (see Table \ref{tab:hardmatch}). Such a low rate of true
matches is a problem for observing campaigns that seek to followup
these sources.  We therefore try to find the source parameters that
maximize the return rate. In particular we wish to locate probable
real matches which are both hard and reddened, which would suggest
that they exist at or beyond the Galactic Center distance.  We present
the statistics for a number of properties and combinations of
properties in Table \ref{tab:hardmatch}.

The most promising property seems to be whether or not a candidate
counterpart has a \textit{J}-band detection. In Table
\ref{tab:hardmatch}, we see that there are $96 \pm 35$ probable true
matches within the set of 1004 NIR single reddened/hard X-ray matches. $51 \pm
13$ of these probable true matches are among the 216 candidate matches
with a \textit{J}-band detection, meaning the remaining $45$ true
matches are among the 788 candidate matches without a \textit{J}-band
detection. Even though the number of probable true matches within the
\textit{J} detected and \textit{J} undetected categories are roughly
the same, the percentage of candidate matches that are real clearly
favors those with \textit{J}-band detections: $27\pm 5 \%$ for
\textit{J} detected versus $6 \pm 4\%$ for \textit{J} undetected.

With this in mind, we break the set of \textit{J} band detected hard
and reddened matches into sub-catagories of \textit{bright/sparse},
\textit{bright/dense}, \textit{faint/sparse} and
\textit{faint/dense}. X-ray sources which are \textit{bright} include
76 single source IR matches and have a 45 $\%$ probability of being
real. If we restrict the error circle sizes of the X-ray sources to
$\sigma_{X} \leq 1.0$, the number of true IR matches drops by 1
source, and the probability of being real rises to 47$\%$ (see
second-to-last row of Table \ref{tab:hardmatch}).  The X-ray faint
sources with J band detections have a lower rate of expected real
matches, 12$\%$. We searched for other parameters that can raise the
percentage of true sources within the X-ray faint category, and we
find that if the sources are matched to $13.0 \leq K_{s} < 14.0~mag$
IR sources (in addition to their being reddened and detected in the
\textit{J}-band), and the X-ray error circles are restricted to
$\sigma_{X} \leq 1.0$, the percentage probability of being real rises
to $40\%$ (see last row of Table \ref{tab:hardmatch}). The $13.0 \leq
K_{s} < 14.0~mag$ criterion is suggested by the color-magnitude
analysis in \S 6.2. In combination, these two categories, shown in the
bottom two lines of Table \ref{tab:hardmatch}, have 98 X-ray/NIR
candidate matches, with $44 \pm 7$ probable real matches.

\begin{table}[ht]
\begin{center}
\caption{Matching statistics for hard X-ray sources matched to a single IR source}
\label{tab:hardmatch}
\begin{tabular}{crrrrrrrrrrr}
\tableline\tableline
Criteria & $N_{obs}$ & $N_{ran}$ & $N_{real}$ & $R=\frac{N_{real}}{N_{obs}}$& $\sigma_{R}$& $\frac{R}{\sigma_{R}}$\\
\tableline
All hard X-ray          &  1223& 1135.0 & 128   & 0.11 & 0.03 & 3.3 \\
Unreddened              &  219 & 193.1  &  33   & 0.15 & 0.06 & 2.4 \\
Reddened                &  1004& 941.9  &  96   & 0.10 & 0.04 & 2.7 \\
With J detection        &  310 & 233.5  &  85   & 0.27 & 0.05 & 5.4 \\
With no J detect.       &  913 & 901.6  &  44   & 0.08 & 0.05 & 1.3 \\
Unred. with J det.      &  94  & 62.5   &  34   & 0.36 & 0.08 & 4.4 \\
Red. with J det.        &  216 & 171.0  &  51   & 0.24 & 0.06 & 3.9 \\
X-ray bright sources    &  324 & 270.2  &  66   & 0.20 & 0.05 & 4.2 \\
X-ray faint sources     &  899 & 864.8  &  82   & 0.09 & 0.06 & 1.5 \\
Dense regions           &  801 & 748.4  &  76   & 0.10 & 0.04 & 2.4 \\
Sparse regions          &  422 & 386.7  &  53   & 0.13 & 0.05 & 2.3 \\
\tableline
Bright,red., w/ J       &  76  & 45.0   &  34   & 0.45 & 0.08 & 5.3 \\      
Faint, red., w/ J       &  136 & 121.9  &  16   & 0.12 & 0.08 & 1.5 \\
Dense, red J            &  166 & 131.6  &  39   & 0.23 & 0.07 & 3.3 \\
sparse red J            &   50 & 39.4   &  12   & 0.25 & 0.12 & 2.0 \\
bright dense red w/ J   &   54 & 34.8   &  22   & 0.41 & 0.10 & 3.9 \\
bright sparse red w/ J  &   23 & 12.3   &  12   & 0.51 & 0.14 & 3.5 \\
faint dense red w/ J    &   112& 96.7   &  17   & 0.15 & 0.09 & 1.6 \\
faint sparse red w/ J   &   27 & 27.0   &  1    & 0.03 & 0.19 & 0.1 \\
\tableline
bright, red. w/ J, $\sigma_{X} \leq 1.0$ &   69 & 39.7   &  32   & 0.47& 0.09  & 5.5\\
faint, red. w/ J, $\sigma_{X} \leq 1.0$, $13.0 \leq K_{s} < 14.0$, &   29 & 17.6   &  12   & 0.40& 0.14  & 2.8\\
\tableline
\end{tabular}
\end{center}
\end{table}

In Figure \ref{fig:colordistributions} we plot the \textit{J-H},
\textit{J-K$_{s}$} and \textit{H-K$_{s}$} histograms for all the
candidate matches followed by the color histograms for the component
of the matches that are expected to be real. The numbers of soft
source matches are shown in blue and the hard source matches are shown
in red.  It is apparent that the ratio of real to candidate sources is
much higher for soft sources than hard sources.  Also, soft sources
with reddened counterparts seem to nearly all be spurious, as are most
hard sources with unreddened counterparts.

\section{Total number of real counterparts in the ISPI catalog}

There are 233 X-ray sources with 2 or more NIR candidate counterparts
in the matched catalog, totalling 527 matches. Of these 233 sources,
84 are X-ray soft (with 185 matches) and 149 are X-ray hard (with 342
matches). We use the rate of real matches, $R=N_{real}/N_{obs}$
derived in \S 4.5 and 4.6 and listed in Tables \ref{tab:softmatch} and
\ref{tab:hardmatch} to estimate the total number of detected true NIR
counterparts within the multiply matched sources.

\subsection{Soft sources}

The real counterpart matching rate for soft X-ray sources is $0.60 \pm
0.032$, as derived from the singly matched sources. When applied to
the 84 multiply matched soft X-ray sources, we expect an additional
$50 \pm 2$ real counterparts.  However, a stronger predictor of real
matches is the presence of a $J$ band detection. Within these 84
sources, 66 have a $J$ detected counterpart. If we apply the rate of
real matches for $J$-detected soft source counterparts of $0.89 \pm
0.017$ we estimate $59 \pm 1$ real matches within the multiply-matched
soft sources. The 18 soft sources with no J band counterpart have no
expected real matches. Thus, if we include singly and multiply matched
soft X-ray sources, we expect there to be $230 + 59 = 289 \pm 13 $
real IR counterparts to soft sources in our ISPI catalog. The fraction
of soft sources with detected counterparts is
$f_{real}=\frac{289}{685} = 0.42 \pm 0.02$.

In the ISPI $17' \times 17'$ area, \cite{mauerhan09} find 324
candidate matches to soft X-ray sources; 200 of these are
unreddened. They do not report on the probable number of true matches
within this field, but they find that within the entire $2^{\circ}
\times 0.8^{\circ}$ field, 890 out of 1007 blue matches are real. If
the ratio of 0.88 holds for the $17' \times 17'$ subfield, then they
discovered $\sim 177$ unreddened real counterparts to soft X-ray
sources in the $17' \times 17'$ ISPI GC field. They report essentially
zero real reddened soft source counterparts making their total
detected fraction of soft X-ray counterparts, $f_{real}\sim
\frac{177}{685}= 0.26$. Our value of $0.42 \pm 0.02$ is larger, which
we expect because our catalog has twice as many detected sources as
the SIRIUS catalog within this area due to our NIR data's greater
depth and resolution.

\subsection{Hard sources}
For hard sources, the rate of real counterparts to number of candidate
counterparts is $0.11 \pm 0.03$, for an estimated total of $16 \pm 5$
within the set of 149 hard sources with multiple matches.

To get a better estimate of the real counterpart content of the
multiply matched sources we break this set into the following
criteria: 

1) \textit{hard} reddened sources with $J$ band detections
($24\%$ real matches) 

2) \textit{hard} unreddened sources with $J$
band detections ($37\%$ real matches) 

3) \textit{hard} reddened
sources \textit{without} $J$ band detections ($6\%$ real matches) 

4) \textit{hard} unreddened sources \textit{without} $J$ band
detections ($0\%$ real matches)

In parentheses is the fraction of real matches to candidate matches
calculated for singly matched sources in Table
\ref{tab:hardmatch}. Since these sources are multiply matched, it is
possible for an X-ray source to have a match for more than one of
criteria 1-4.

To reduce the complexity we note that the real match probability for
X-ray sources with $J$ band detected matches is much higher than for
X-ray sources without $J$ band detected matches. Therefore, we assume
that if one of these multiply matched hard sources has a $J$ detected
counterpart, the possibility that any additional matche without a $J$
detection is the real counterpart is negligible.

There are 65 multiply matched hard X-ray sources with at least one $J$
detected match. Four(4) have both unreddened and reddened $J$ band
detected matches, 28 have reddened $J$ band detected matches and 33
have unreddened $J$ band detected matches. For the 4 sources with both
reddened and unreddened matches, we estimate $0.4 \pm 0.1$ real
reddened matches and $0.9 \pm 0.2$ real unreddened matches. The 28
X-ray sources with reddened $J$ band matches should contain $7 \pm 2$
real matches and the 33 X-ray sources with unreddened $J$ detected
matches should have $12 \pm 3$ real matches.

The remaining sources 84 sources do not have $J$ detections. Since we
expect $0 \%$ of the unreddened sources without $J$ detections to have
matches, we only consider the subset of 76 X-ray sources with reddened
candidate matches. Using the real match probability of $0.06$, we
estimate an additional $4 \pm 3$ real reddened matches to hard X-ray
sources from this set of sources.

In total, we find that there should be $24 \pm 4$ additional real
matches to the 149 multiply matched hard X-ray sources. $11 \pm 3$ are
reddened and $13 \pm 3$ are unreddened.

The total number of real, hard X-ray source counterparts for
singly and multiply matched data is $129 + 24 =153 \pm 39$
sources. $107 \pm 37$ sources are hard and reddened, and $44 \pm 14$
are hard and unreddened.

This amounts to a fraction of $f_{real}=\frac{153}{3583} = 0.04 \pm
0.01$ hard X-ray sources with detected counterparts. \cite{mauerhan09}
found a rate of $0.058 \pm 0.015$ with data that is slightly shallower
in depth but that spans a larger area around the GC; nevertheless, our
fraction is still consistent with \cite{mauerhan09} within the errors.

For a fairer comparison, we use their statistics on the $8'$ radius
region around Sgr A$^{*}$ ($\alpha=266.41726$,
$\delta=-29.00798$). They calculate the number of real hard reddened
counterparts within this region to be $46.2 \pm 23.1$ sources out of
617 candidate matches.

Our total for this region is $73 \pm 33$ sources out of 1018 X-ray
sources with candidate matches. This fraction is consistent with
\cite{mauerhan09} and our greater total is likely due to the greater
depth and resolution of the ISPI imaging with respect to the SIRIUS
data.

\section{CMD positions of probable real matches}

We constructed an infrared color-magnitude diagram for probable real
matches to X-ray sources. We divided the X-ray sources into soft and
hard and then used the matching statistics within the
$H-K_{s}$/$K_{s}$ CMD with equations A3 and A4 to locate
color-magnitude regions where there are excess matches to the aligned
X-ray catalog. We calculate the excess in matches using bins of 1
magnitude in $K_{s}$ and 1 magnitude in $H-K_{s}$, ranging from
$0.0~<~H-K_{s}~<~4.0~mag$ and $8.0~<~K_{s}<~17.0~mag$. This approach
has the advantage that it does not exclude any region of the CMD a
priori, as would happen if we forced a $J$ detection or a color cut.

In Figures \ref{fig:softcmd} and \ref{fig:hardcmd} we show the CMD of
the entire ISPI catalog overplotted by boxes which are color-coded to
represent the number of probable true counterparts within a CMD
bin. Here we only include X-ray sources with a single IR match. The
box colors are black, yellow, green and blue, and correspond to $0$,
$~1-5$,$~5-9$, and $~>9$ real matches, respectively. Bins with less
than a $2\sigma$ significance
($\frac{N_{real,1}}{\sigma_{N_real,1}}~<~2.0)$ are also marked in
black.

\subsection{CMD of soft source counterparts}

\begin{figure}
\centering
\subfigure[] 
{
    \label{fig:softcmd:a}
    \includegraphics[width=9cm]{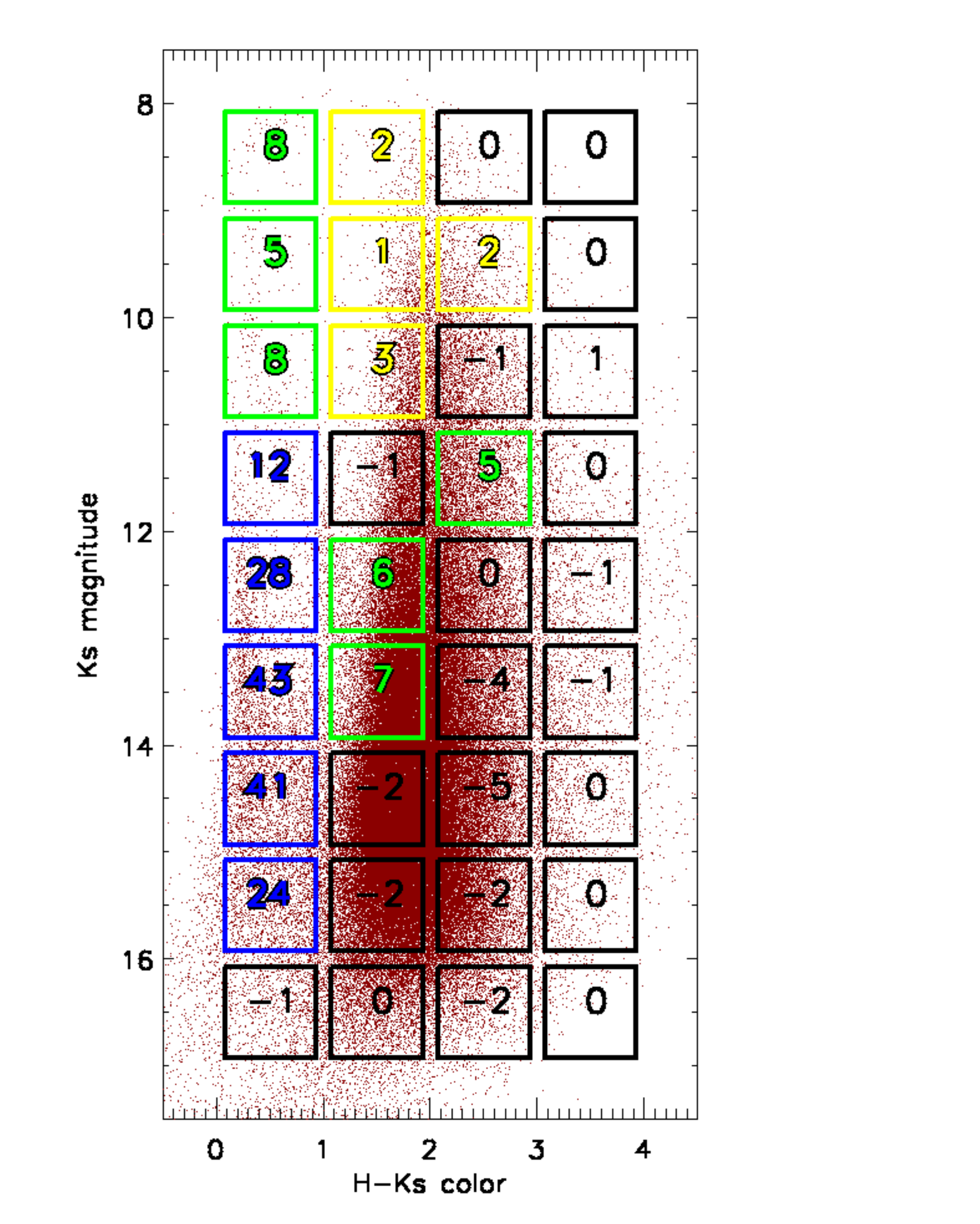}
}
\hspace{-3cm}
\subfigure[] 
{
    \label{fig:softcmd:b}
    \includegraphics[width=9cm]{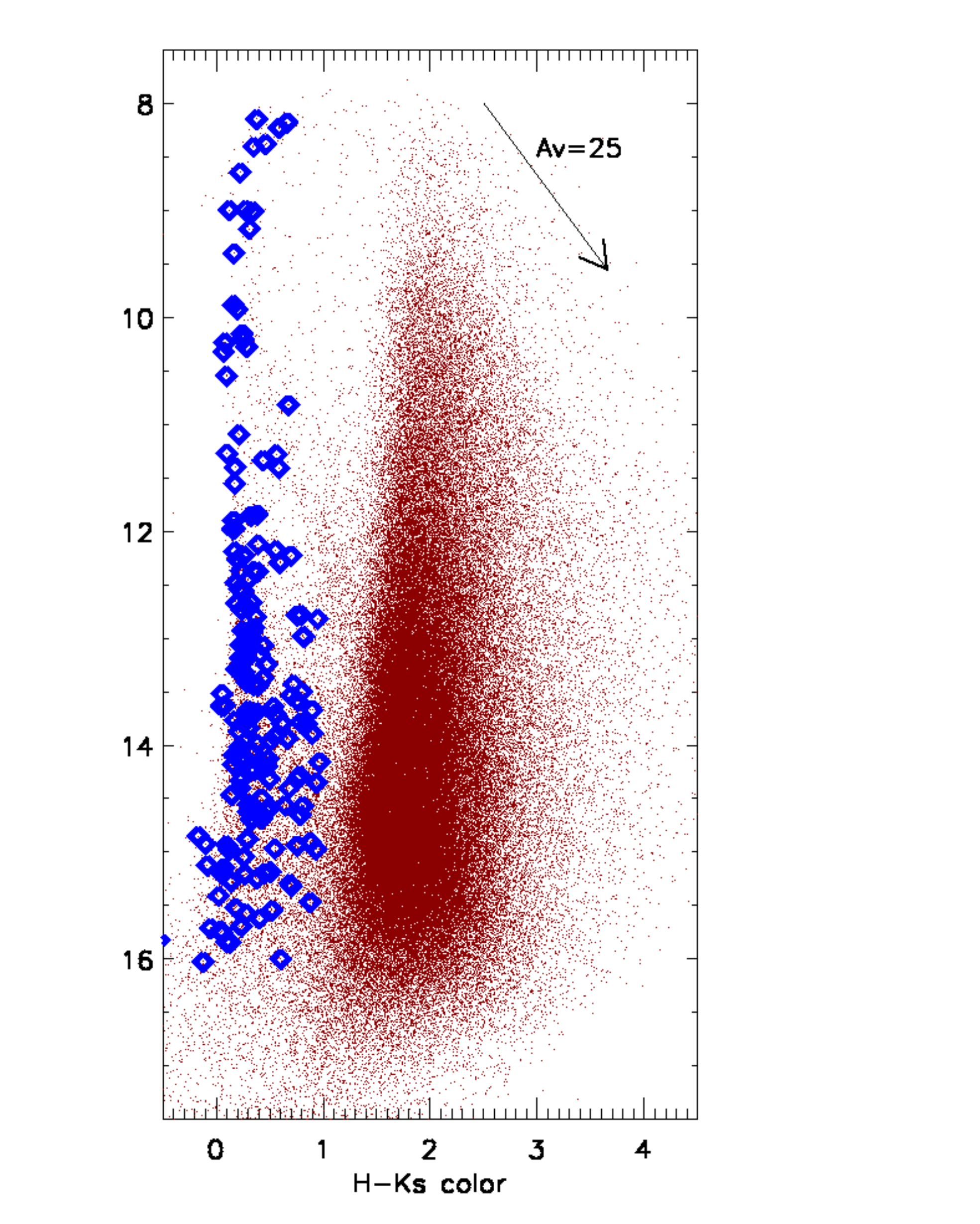}
}
\caption{(left) $H-K_{s}$, K$_{s}$ color magnitude diagram of all ISPI
  sources, in red, with the positions of probable true counterparts to
  X-ray soft sources shown by the colored boxes. The color coding is
  yellow: 1-5 sources, green: 5-9 sources, and blue: 9 or more 
  sources. Black boxes contain counts with lower than $2~\sigma$
  significance and are consistent with 0 real counterparts. The boxes
  are undersized for clarity.  (right) $H-K_{s}$, $K_{s}$ color
  magnitude diagram of all ISPI sources, in dark red. Overplotted in
  blue are a set of 242 soft X-ray matches with J band detections and
  low reddening. These 242 IR sources should have a 96\% likelihood of
  being the true counterparts to their matched X-ray sources. See
  \S 4.5 for details.}
\label{fig:softcmd} 
\end{figure}

In Figure \ref{fig:softcmd} we see that the real soft sources are
mostly confined to $H-K_{s} < 1~mag$, and $8.0 < K_{s} < 16.0$ mag,
with a small number in the bright but more reddened CMD bins ($H-K_{s}
> 1~mag$, and $9.0 < K_{s} < 12.0$ mag).  The number of unreddened
counterparts are consistent with the contribution expected from
coronally active stars in the field \citep{mauerhan09,muno09}. The
fact that the X-ray spectra appear soft suggests that these sources
lie within 4 kpc of the sun \citep{mauerhan09}. Since the stars can be
arbitrarily closer than this distance, and dwarf stars have a wide
dispersion in magnitudes, it is difficult to place further constraints
on the population of unreddened IR counterparts to soft X-ray sources
without spectroscopy.

The real counterparts to soft sources with $H-K_{s} > 1~mag$ are more
mysterious since the high reddening implied by the IR color should
also make the X-ray component appear hard. However, \cite{mauerhan09}
do note that for sources near the hard/soft threshhold, the
uncertainty could cause a hard source to be labeled as a soft source,
and vice versa. Another possibility is that these systems may have
internal absorption that affects the IR component more than the X-ray
component. Finally, for some sources that are particularly strong
soft X-ray emitters, the large extinction may not soften it enough to
change its hard/soft designation. One such source in our sample
(CXO174550.6-285919), is nominally soft and reddened but spectroscopic
follow-up found that it was a Wolf-Rayet binary in the Galactic Center,
with a spectral class WN6b \citep{mauerhan10}.

In \S 4.5 we found that $96 \%$ of the 242 singly matched soft X-ray
source counterparts with unreddened colors and $J$ detections are
likely to be authentic. We plot their positions over the ISPI CMD in
the right panel of Figure \ref{fig:softcmd}.  However, because we
used unreddened colors and a $J$ band detection as a criteria in order
to enhance the probability, any real reddened matches to soft X-ray
sources will not represented in this figure.

\subsection{CMD of hard source counterparts}

\begin{figure}
\centering
\subfigure[] 
{
    \label{fig:hardcmd:a}
    \includegraphics[width=9cm]{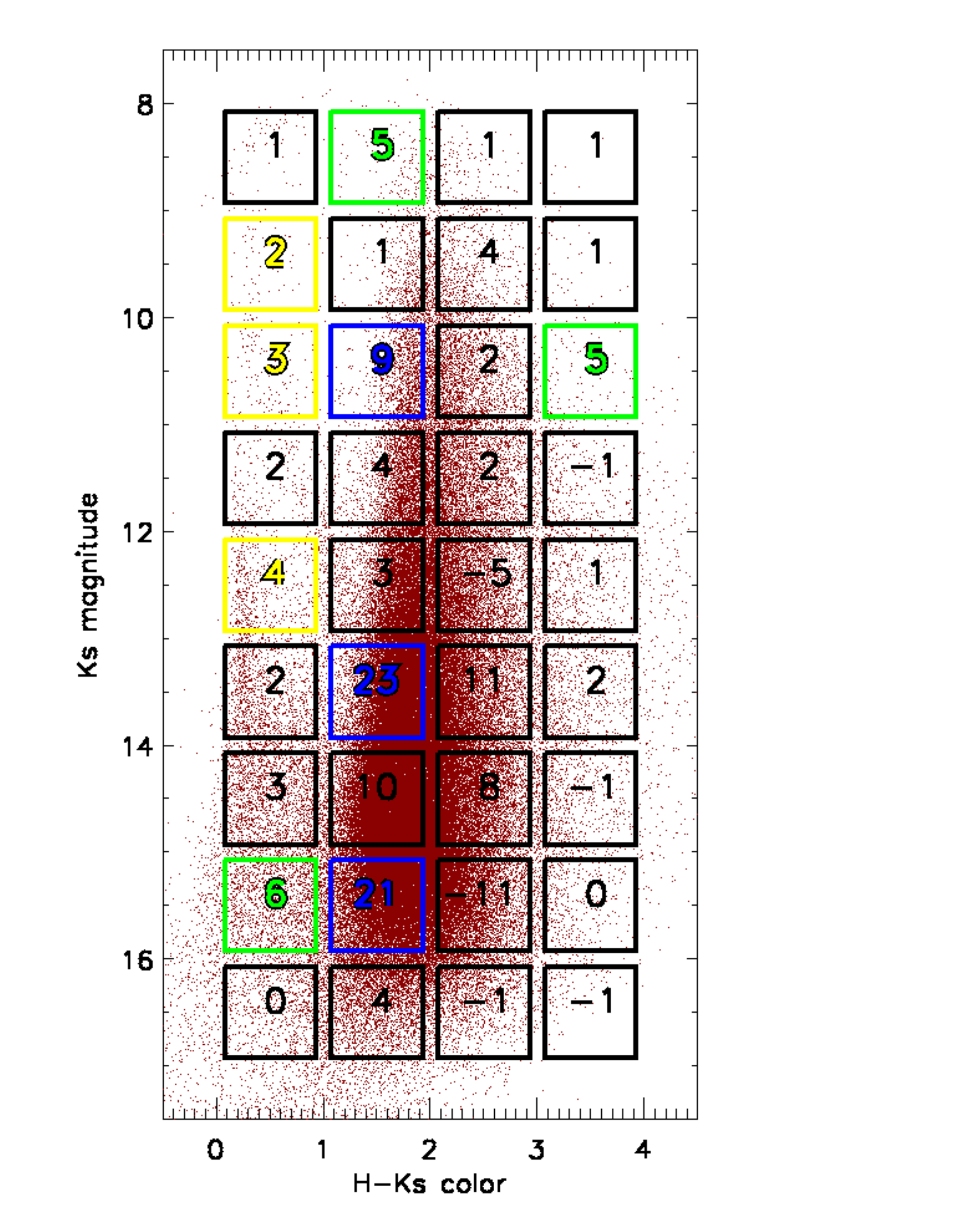}
}
\hspace{-3cm}
\subfigure[] 
{
    \label{fig:hardcmd:b}
    \includegraphics[width=9cm]{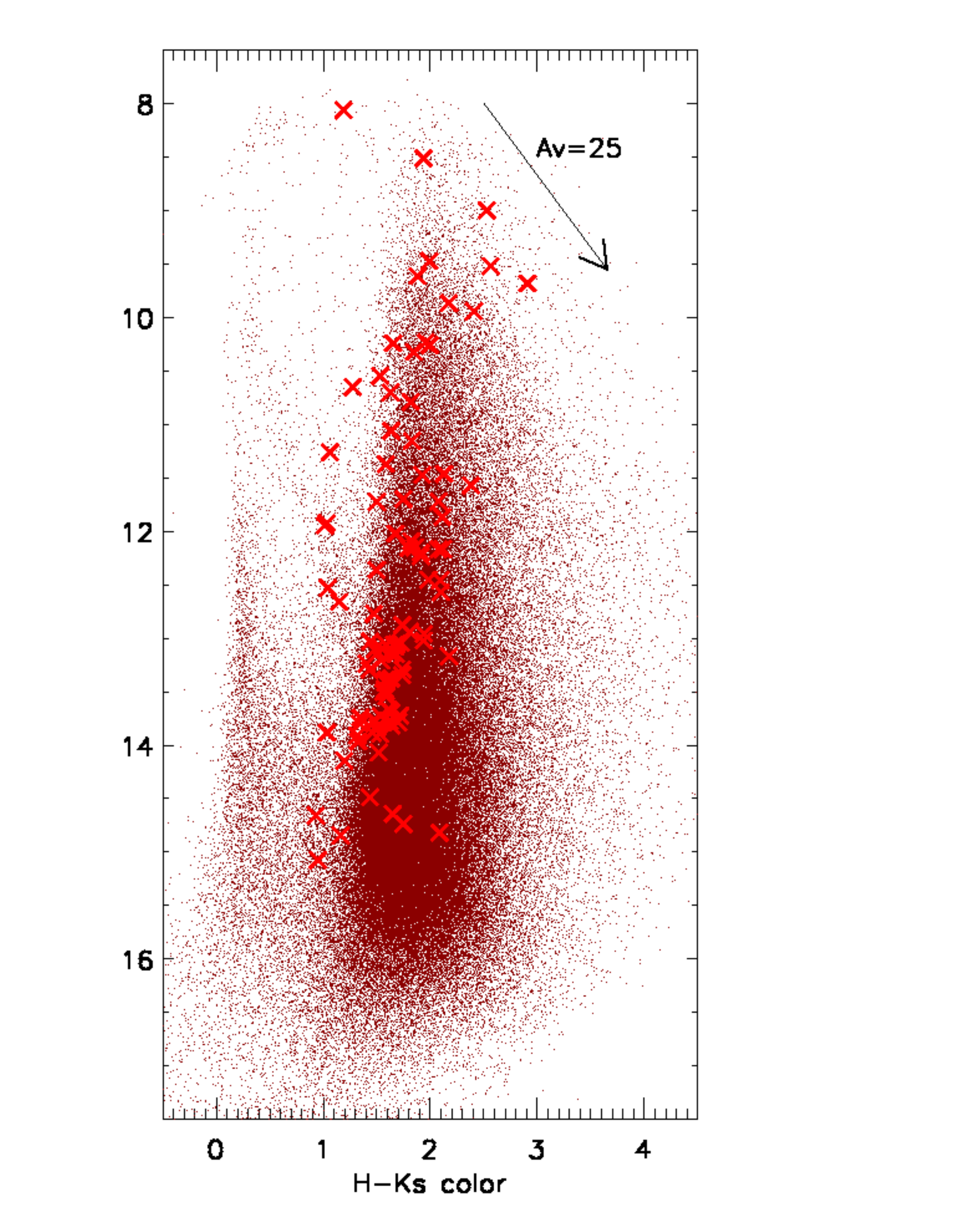}
}
\caption{(left) $H-K_{s}$, $K_{s}$ color magnitude diagram of all ISPI
  sources, in red, with the positions of probable true counterparts to
  hard X-ray sources shown by the colored boxes. The color coding is
  yellow: 1-5 sources, green: 5-9 sources, and blue: 9 or more
  sources. Black boxes contain counts with lower than $2~\sigma$
  significance and are consistent with 0 real counterparts. The boxes
  are undersized for clarity. (right) $H-K_{s}$, $K_{s}$ color
  magnitude diagram of all ISPI sources, in dark red. Overplotted are
  a set of 98 hard X-ray matches with reddened colors, J band
  detections, small positional errors ($\sigma_{X}\leq 1.0"$) and are
  either X-ray bright or X-ray faint with $13\leq K_{s}<14~mag$. These
  98 IR sources should have a 45\% likelihood of being true
  counterparts to their matched X-ray sources. See \S4.6 for details.}
\label{fig:hardcmd} 
\end{figure}

The hard X-ray source CMD has three bins with a 2$\sigma$ detection of
9 or more real matches (the blue boxes in Figure \ref{fig:hardcmd}).
All of these CMD bins reside in the $1 < H-K_{s} < 2~mag$ column of
the CMD. The magnitudes at which these appear are not contiguous, with
$9 \pm 3$ sources at $10 < K_{s} <11~mag$, $23 \pm 9$ sources at $13
< K_{s} <14~mag$ and $21 \pm 9$ sources at $15 < K_{s} <16~mag$. There
are also a handful of true counterpart detections in the unreddened
bins dispersed between $10 < K_{s} <16~mag$ as well as $5 \pm 1$
likely true counterparts at $10 < K_{s} <11~mag$ and $3 <H-K_{s}< 4~
mag$. 

In the right panel of Figure \ref{fig:hardcmd} we show the CMD
positions of the 98 hard, reddened matches with a $J$-band detection,
of which $\sim 45 \%$ should be real as discussed in \S 4.6. These
sources represent the highest proportion of reddened authentic
counterparts to hard X-ray sources that we could locate in the matched
catalog. Even so, half of these counterparts are spurious. Since the
criterion of a $J$ detection was used to enhance the real counterpart
probability, the lower right CMD is excluded due to flux sensitivity.

The unknown distance, unknown extinction and unknown intrinsic
$(H-K_{s})_{0}$ color make interpretation of the hard counterpart CMD
complicated. Nevertheless, a useful constraint on the $M_{K_{s}}$
magnitudes can be found by considering ranges of the uncertainty in
these parameters.

\cite{muno09} argue that the majority of hard X-ray sources must be
farther than 4kpc from the Sun, because of the extinction column
required to make soft sources appear as hard. We adopt this value as
the minimum distance for hard X-ray sources. For the maximum distance,
we use a value of 8kpc, a reasonable value for the distance to the
Galactic Center. There is the possibility of observing bright IR
counterparts on the far side of the bulge (with distance $> 8kpc$).
However, the strongest candidate CMD bins lie in the $1 < H-K_{s} <
2~mag$ range, which is mostly to the blue side of the modal value of
$H-K_{s}=1.8~mag$ for the ISPI catalog, implying that they may lie on
the near side of the GC.

The intrinsic colors for main sequence, giant and supergiant stars
range from $-0.06 < (H-K)_{0} < 0.36~mag$, and $-0.06 < (H-K)_{0} <
0.1~mag$ for all spectral types except for M \citep{cox00}. We use
this dispersion in intrinsic color and the $1 < H-K_{s} < 2~mag$
limits of the color bin to calculate the range in absolute $K_{s}$
extinction, $A_{K_{s}}$. Using extinction ratios for the Galactic
Center of $A_{V}:A_{J}:A_{H}:A_{K_{s}}=1.00:0.188:0.108:0.062$
\citep{nishiyama08} we find that sources with $1 < H-K_{s} < 2~mag$
are experiencing $0.86 < A_{K_{s}} < 2.78~mag$ of extinction.

We estimate the absolute $M_{K_{s}}$ magnitude range for the blue bins
in Figure \ref{fig:hardcmd} using equation 10 with the magnitude
ranges of the bins and the estimated $A_{K_{s}}$ range.
\begin{equation}
M_{K_{s}}=K_{s}-5log(d)+5-A_{K_{s}}
\end{equation}
The resulting limits on $M_{K_{s}}$ are listed in Table
\ref{tab:magrange}.

The ranges of possible $M_{K_{s}}$ are broad: each 1 magnitude bin in
apparent magnitude can accommodate an interval of $\sim 3~mag$ in
absolute magnitude for certain distances and extinctions.  But we can
still suggest certain classes of object that meet the criteria for
these probable real sources. 

The systems in the $10<K_{s}<11~mag$ bin have a possible range of
$-2.9<M_{K}<-7.3~mag$. The matched catalog contains 20 candidate
matches in this bin, and 4 of these have had published spectroscopic
followup. \cite{mauerhan10} observed $K$ band spectra of the NIR
counterparts to CXO174532.7-285617, CXO174536.1-285638,
CXO174555.3-285126, CXO174617.0-285131 which all lie within the
$10<K_{s}<11~mag$, $1<H-K_{s}<2~mag$ bin and are X-ray hard. They
found spectral types of O4-6I, WN8-9h, WN5-6b, O6If+
respectively. Supergiants of all spectral types are brighter than
$M_{K} \sim -6~mag$, as are Wolf Rayet stars and O main sequence stars
\citep{cox00,carroll96}. At the GC extinction levels, Wolf Rayet stars
and single O stars are usually too soft in X-rays to be
detected. However, if they exist in a binary, with either an accreting
compact object or a star with high mass loss, then this can create a
wind shock zone energetic enough to generate hard X-rays.

The sources in the $13<K_{s}<14~mag$, $1< H-K_{s}< 2~mag$ bin should
come from stars with $0.1<M_{K}<-4.3~mag$ and the
$15<K_{s}<16~mag$, $1< H-K_{s}< 2~mag$ bin sources correspond to
$2.1~<~M_{K}~<~-2.3~mag$. Both ranges overlap well with giant stars in
luminosity class III, which have $M_{K_{s}}= -0.75~mag$ for G types
and get monotonically brighter to $M_{K_{s}}= -5~mag$ for mid M type
giants \citep{carroll96,cox00}. These stars could produce the
requisite hard X-rays if they belonged to an accreting binary system,
such as a LMXB or a symbiotic binary (containing a RGB star and a
white dwarf). Most symbiotic stars have soft X-ray spectra but recent
evidence suggests that symbiotic stars with a highly magnetized white
dwarf have hard X-ray spectra \citep{luna07}.

B0-B8 stars on the main sequence have $0 < M_{K_{s}}< -3~mag$, which
also makes them good candidates for the $13~<~K_{s}~<14~mag$ and
$15~<K_{s}<~16~mag$ bins. B star HMXBs are known to have hard spectra
and would be expected to exist in regions where there has been recent
star formation. This would be consistent with the known existence of
young massive stars the GC. 

\begin{table}
\begin{center}
\caption{Ranges of $M_{K_{s}}$ possible for real sources in CMD bins from Figure \ref{fig:hardcmd}. The color range is $1.0<(H-K_{s})< 2.0$ for all three bins.}
\label{tab:magrange}
\begin{tabular}{crrrrrrrrrrr}
\tableline\tableline
Apparent magnitude range & $M_{K_{s}}$ for $d=4kpc$ & $M_{K_{s}}$ for $d=8 kpc$ \\
$10<K_{s}<11$            & $ -5.8<M_{K_{s}}<-2.9 $  &  $ -7.3 <M_{K_{s}}<-4.4  $ \\
$13<K_{s}<14$            & $ -2.8<M_{K_{s}}< 0.1 $  &  $ -4.3 <M_{K_{s}}<-1.4  $ \\
$15<K_{s}<16$            & $ -0.8<M_{K_{s}}<~2.1 $  &  $ -2.3 <M_{K_{s}}<~0.6  $ \\
\tableline
\tableline
\end{tabular}
\end{center}
\end{table}

\section{Conclusions}

We cross-correlated a new $JHK_{s}$ catalog of the central $17'\times
17'$ of the Galaxy with a catalog of 9017 X-ray point sources observed
with \textit{Chandra}. Our catalog of astrometrically matched
candidate counterparts contains 2137 IR matches to 1843 X-ray
sources. Through Monte Carlo simulations of the matching precedure we
determine that the candidate matches contain $289 \pm 13$ true matches
to soft X-ray sources and $154 \pm 39$ true matches to hard X-ray
sources.  The soft X-ray sources are likely to be a mix of coronally
active dwarf stars within 4 kpc, and we find that, by choosing the
candidate singly matched 242 IR/X-ray candidate counterparts with
unreddened infrared colors and $J$ band detections, there will be $232
\pm 2$ real NIR counterparts.

Hard, reddened sources, which other authors have suggested to be a mix
of HMXBs, colliding wind binaries and other rare systems in the
Galactic Center vicinity, should have $110 \pm 40$ real infrared
matches in our catalog, out of 1372 candidates. We calculate the
difference in CMD diagrams between candidate matches and simulated
spurious matches and locate 3 distinct regions of the
$H-K_{s}$/$K_{s}$ CMD with a significant number of probable real
counterparts. Hidden behind between $0.86 < A_{K_{s}} < 2.78~mag$ of
extinction, some of these sources should have absolute magnitude
ranges compatable with Wolf-Rayet stars and supergiants, while the
fainter two bins are more likely to be main sequence B star HMXBs or
RGB stars with a compact accreting companion (possibly a neutron star
or black hole LMXB or as a symbiotic star with a magnetic white dwarf).

The low probability ($\sim 10\%$) of finding real matches in the set
of hard match candidates is a challenge for spectroscopic follow-up
campaigns.  We find sets of properties which enhance the probability
of obtaining real matches to X-ray hard, IR-reddened sources. We find
that by restricting the hard, reddened candidate matches to those that
are X-ray bright, with positional errors $\sigma_{X} \leq 1.0$, and
that are matched to an IR source with a J band detection, or from
X-ray faint sources with small positional errors ($\sigma_{X} \leq
1.0$) with $13.0 \leq K_{s} < 14.0~mag$ and a J band detection, one is
likely to find $44 \pm 7$ real IR counterparts out of a total of 98
candidate counterparts. These results will be used for target
selection and slit design to maximize the return of upcoming NIR
spectroscopic campaigns, such as the FLAMINGOS-2 Galactic Center
Survey \citep{eiken08}.

\appendix

\section{Appendix}

In section \S 4.2 we derived equations that estimate the number of
singly matched X-ray sources with real NIR counterparts for X-ray
sources with a given set of X-ray properties. We wish to extend this
analysis to ask the question, how many real NIR counterparts are there
for a given set of X-ray properties \textit{and NIR properties}? 

In this case, there is a bias that occurs when a real NIR
counterpart is randomized to match a NIR source with the opposite NIR
properties.

We begin by defining $P$ to be a set of NIR properties (such as being
within a ($H-K_{s}$) color interval or having a $J$ band detection).

$N_{obs,1}$ is the number of singly matched X-ray sources in the
aligned X-ray catalog whose counterparts fall within $P$. $N_{obs,2}$
is the number singly matched X-ray sources whose counterparts fall
within the \textit{complement} of $P$.

$N_{obs,1}$ and $N_{obs,2}$ consist of both real and spurious NIR
matches.
\begin{equation}
N_{obs,1}= N_{real,1}+N_{spur,1}
\end{equation}
and 
\begin{equation}
N_{obs,2}= N_{real,2}+N_{spur,2}
\end{equation}

We define $N_{ran,1}$ and $N_{ran,2}$ to be the average number of
singly matched X-ray sources to the randomized X-ray catalogs whose
counterparts are within $P$ and the \textit{complement} of $P$,
respectively.

As in \S 4.2, we assume the spurious match rate from the aligned X-ray
catalog is the same as for the randomized catalog, and we find $N_{ran,1}$
and $N_{ran,2}$ to be:
\begin{equation}
N_{ran,1}= A \times \frac{N_{spur,1}}{A-N_{real,1}-N_{real,2}}
\end{equation}
and for spurious matches to sources without property P:
\begin{equation}
N_{ran,2}= A \times \frac{N_{spur,2}}{A-N_{real,1}-N_{real,2}}
\end{equation}

The result is four linear equations and four unknown variables, $N_{real,1}$,$N_{spur,1}$,
$N_{real,2}$ and $N_{spur,2}$. Solving via Gaussian elimination gives:
\begin{equation}
N_{spur,1}= \frac{N_{ran,1}(A-N_{obs,1}-N_{obs,2})}{A-N_{ran,1}-N_{ran,2}}
\end{equation}
\begin{equation}
N_{real,1}= N_{obs,1}-\frac{N_{ran,1}(A-N_{obs,1}-N_{obs,2})}{A-N_{ran,1}-N_{ran,2}}
\end{equation}
\begin{equation}
N_{spur,2}=\frac{N_{ran,2}(A-N_{obs,1}-N_{obs,2})}{A-N_{ran,1}-N_{ran,2}}
\end{equation}
\begin{equation}
N_{real,2}=N_{obs,2}-\frac{N_{ran,2}(A-N_{obs,1}-N_{obs,2})}{A-N_{ran,1}-N_{ran,2}}
\end{equation}
The errors in the expected number of real counterparts follow from
the standard rules of error propagation.
\begin{equation}
  \sigma_{N_{real,1}}=\sigma_{N_{spur,1}}=\sqrt{(\frac{(A-N_{ran,2})(A-N_{obs,1}-N_{obs,2})}{(A-N_{ran,1}-N_{ran,2})^{2}})^{2}\sigma_{N_{ran,1}}^{2}+(\frac{N_{ran,1}(A-N_{obs,1}-N_{obs,2})}{(A-N_{ran,1}-N_{ran,2})^{2}})^{2}\sigma_{N_{ran,2}}^{2}}
\end{equation}
\begin{equation}
  \sigma_{N_{real,2}}=\sigma_{N_{spur,2}}=\sqrt{(\frac{(A-N_{ran,1})(A-N_{obs,1}-N_{obs,2})}{(A-N_{ran,1}-N_{ran,2})^{2}})^{2}\sigma_{N_{ran,1}}^{2}+(\frac{N_{ran,2}(A-N_{obs,1}-N_{obs,2})}{(A-N_{ran,1}-N_{ran,2})^{2}})^{2}\sigma_{N_{ran,2}}^{2}}
\end{equation}

The variable definitions for this derivation are collected in Table
\ref{tab:vardefs2}.

\begin{table}[ht]
\begin{center}
  \caption{Variable definitions for calculating numbers of real
    counterparts to X-ray sources with a given IR counterpart
    property}
\label{tab:vardefs2}
\small
\begin{tabular}{crrrrrrrrrrr}
\tableline\tableline
Variable name & Meaning\\
\tableline
$A$              &  total num. of X-ray sources with a given set of X-ray properties  \\
$N_{obs,1}$        &  num. of singly matched X-ray sources from aligned X-ray catalog w/ P\\ 
$N_{obs,2}$       &  num. of singly matched X-ray sources from aligned X-ray catalog w/o P\\ 
$N_{ran,1}$        & mean num. of singly matched X-ray sources from rand. X-ray catalogs w/ P\\
$N_{ran,2}$        & mean num. of singly matched X-ray sources from rand. X-ray catalogs w/o P\\
$\sigma_{N_{ran,1}}$& standard deviation of $N_{ran,1}$ \\
$\sigma_{N_{ran,2}}$& standard deviation of $N_{ran,2}$ \\
$N_{real,1}$         & num. of \textit{real} IR counterparts within N matches w/ P \\
$N_{spur,1}$         & num. of \textit{spurious} IR counterparts within N matches w/ P \\
$N_{real,2}$         & num. of \textit{real} IR counterparts within N matches w/o P \\
$N_{spur,2}$         & num. of \textit{spurious} IR counterparts within N matches w/o  P \\
$\sigma_{N_{real,1}}$& $1\sigma$ error in value of $N_{real,1}$\\
$\sigma_{N_{spur,1}}$& $1\sigma$ error in value of $N_{spur,1}$\\
$\sigma_{N_{real,2}}$& $1\sigma$ error in value of $N_{real,2}$\\
$\sigma_{N_{spur,2}}$& $1\sigma$ error in value of $N_{spur,2}$\\
\tableline
\end{tabular}
\end{center}
\end{table}

\clearpage

\begin{deluxetable}{ccrrrrrrrrcrl}
\tabletypesize{\scriptsize}
\rotate
\tablecaption{Sample Data of ISPI/Chandra matched sources catalog}
\tablewidth{690.71169pt}
\tablehead{
\colhead{X-ray ID} & \colhead{X-ray RA (deg)} & \colhead{X-ray DEC (deg)} & \colhead{$\sigma_{X}$ (")} & \colhead{Source type} &
\colhead{ISPI RA (deg)} & \colhead{ISPI DEC (deg)} & \colhead{J} &
\colhead{J error} & \colhead{H} &
\colhead{H error} & \colhead{K$_{s}$} & \colhead{K$_{s}$ error}
}
\startdata
 174457.1-285740& 266.23813 & -28.96121 & 1.8& hard & 266.238118 & -28.961     & 14.54 & 0.034& 12.98& 0.037& 11.60 & 0.043\\
 174457.4-285622& 266.23941 & -28.93967 & 2.0& hard & 266.239377 & -28.940386  & 18.09 & 0.069& 14.01& 0.043& 12.74 & 0.051\\
 -		& -	    & -         & -  & -    & 266.239183 & -28.939192  & 17.81 & 0.064& 14.01& 0.043& 11.89 & 0.044\\
 174459.9-290324& 266.24982 & -29.05683 & 2.0& hard & 266.249972 & -29.056337  & 13.66 & 0.031& 13.06& 0.037& 11.20 & 0.041\\
 174459.9-290538& 266.24994 & -29.09415 & 1.5& soft & 266.250792 & -29.094179  & 15.57 & 0.039& 12.18& 0.034& 11.70 & 0.043\\
 174500.2-290057& 266.25113 & -29.01598 & 2.6& hard & 266.25107  & -29.015537  & 14.87 & 0.035& 12.52& 0.035& 10.83 & 0.039\\
 174501.9-285719& 266.25827 & -28.95553 & 1.8& hard & 266.258535 & -28.955952  & 14.44 & 0.033& 13.02& 0.037& 12.40 & 0.048\\
 174502.2-285749& 266.25946 & -28.96381 & 1.0& hard & 266.259295 & -28.963779  & 14.11 & 0.032& 11.75& 0.033& 10.41 & 0.038\\
 174502.4-290205& 266.26007 & -29.03492 & 1.1& hard & 266.259337 & -29.03475   & 13.95 & 0.032& 11.89& 0.033&  9.83 & 0.037\\
 -		& -         & -         & -  & -    & 266.260048 & -29.034872  & 13.84 & 0.031& 11.93& 0.033&  9.98 & 0.037\\
 174502.4-290453& 266.26039 & -29.0816  & 1.4& soft & 266.260307 & -29.081633  & 15.47 & 0.038& 13.04& 0.037& 11.31 & 0.041\\
 174502.8-290429& 266.26198 & -29.0748  & 1.3& hard & 266.261948 & -29.074818  & 13.41 & 0.030& 12.93& 0.037& 12.76 & 0.052\\
 174504.2-290410& 266.26764 & -29.06977 & 1.5& hard & 266.268021 & -29.070093  & 15.42 & 0.038& 11.62& 0.033&  9.75 & 0.036\\
 174504.2-290610& 266.26841 & -29.10258 & 1.9& soft & 266.268758 & -29.102386  & 16.39 & 0.046& 12.31& 0.034& 10.78 & 0.039\\
 
\enddata
\end{deluxetable}

\clearpage

\end{document}